\documentclass[11pt,letterpaper]{JHEP3}

\usepackage{epsfig} 
\usepackage{amssymb} 
\usepackage{amsmath} 
\usepackage{amsfonts} 
\usepackage{slashed} 
\usepackage{axodraw} 
\usepackage{cite}

\def\ltap{\ \raise.3ex\hbox{$<$\kern-.75em\lower1ex\hbox{$\sim$}}\ }
\def\gtap{\ \raise.3ex\hbox{$>$\kern-.75em\lower1ex\hbox{$\sim$}}\ }
\def\eg{{\it e.g.}}
\def\ie{{\it i.e.}}
\def\be{\begin{equation}}
\def\ee{\end{equation}}
\def\bea{\begin{eqnarray}}
\def\eea{\end{eqnarray}}
\newcommand{\beq}{\begin{eqnarray}}% can be used as {equation} or {eqnarray}
\newcommand{\eeq}{\end{eqnarray}}
\newcommand{\nn}{\nonumber}

\newcommand{\vev}[1]{ \left\langle {#1} \right\rangle }
\newcommand{\bra}[1]{ \langle {#1} | }
\newcommand{\ket}[1]{ | {#1} \rangle }
\newcommand{\gev}{{\rm GeV}}
\newcommand{\met}{\slashed {E}_{T}}
 
\newcommand{\eq}[1]{(\ref{#1})}  
\newcommand{\z}{\frac{m_f^2}{m_\chi^2}}
\newcommand{\vr}{v_\text{rel}}
\newcommand{\sfrac}[2]{#1/#2}
\renewcommand{\vec}[1]{\mbox{\boldmath $#1$}}

\newcommand{\gm}{\gamma^\mu}

\newcommand{\gf}{\gamma_5}
\DeclareMathOperator{\su}{SU}

\title{LHC and Dark Matter Signals of Improved Naturalness}
\author{R.~Enberg$^{a,b}$, P.~J.~Fox$^b$, L.~J.~Hall$^b$, A.~Y.~Papaioannou$^b$, M.~Papucci$^b$\\
\\
$^a$ Department of Physics, University of Arizona, Tucson, AZ 85721\\
$^b$ Theoretical Physics Group, Lawrence Berkeley National Laboratory, and\\
Department of Physics, University of California, Berkeley, CA 94720}

\abstract{The Standard Model Higgs suffers from the hierarchy problem, 
typically implying new states within the reach of the LHC.  
If the Higgs is very heavy ($\sim 500$~GeV) the states that 
cutoff the quadratic divergence may be beyond the reach of the LHC.  
However, in this case precision electroweak data require the Standard 
Model to be augmented with new states at the electroweak scale.  
We study a very simple model, with no new colored states, that allows 
a heavy Higgs whilst remaining consistent with experiments, 
and yielding the correct dark matter abundance.  We investigate the 
possibilities for its discovery at the LHC and future dark matter 
detection experiments.}

\preprint{LBNL-62748\\UCB-PTH-07/10}

\begin{document}

%============================================================================== 
\section{Introduction}
%============================================================================== 

The Large Hadron Collider will finally decide the question of whether
the Higgs boson exists and, if it does, it will provide an accurate
measurement of its mass.  In recent decades, the Standard Model has 
been so successful in describing all data in particle physics, that it 
is tempting to use it to predict what the LHC will find.  Precision
electroweak data, including the lower value for the top quark mass
from recent running of the Fermilab Tevatron, leads to a prediction
for the Higgs boson mass of $m_H = 76^{+33}_{-24}$ GeV at 68\% confidence level~\cite{lepewwg}.
Since the LEP direct search limit is 114 GeV, the Standard Model predicts
that LHC will discover the Higgs with a mass not far from this current
direct limit\footnote{Combining precision electroweak data with the LEP direct searches one gets an upper bound of $m_{H} < 182$ GeV at 95\% confidence level.}.

The Standard Model is an effective theory valid only up to some energy cutoff $\Lambda$ which, if symmetry principles govern fundamental physics, is possibly well below the Planck scale.   
Requiring that it provides a natural description of electroweak
symmetry breaking, \ie{} that the Higgs mass does not receive dominant
loop corrections from the cutoff scale, leads to an estimate of the
maximum cutoff scale:\  $\Lambda_{max} \approx 3.7\, m_H$
from a radiative loop involving the top quark~\cite{Barbieri:2006dq,Barbieri:2005ri}. 
The prediction of a light Higgs is
therefore very exciting, implying that the Standard Model will be
incorporated into a new theory in the energy domain accessible to the
LHC, which would be expected to discover this new ``canceling physics''.
In the case of the top quark this is typically expected to be provided by
new colored states lighter than 1 TeV. 

However, while these arguments are plausible, they should not be
believed. If the Standard Model Higgs sector breaks down at an energy
scale as low as 400 GeV, why do we expect it to be correct at, say,
200 GeV?  A frequent answer to this is that a light Higgs neatly accounts for
the observed sizes of the precision electroweak vacuum polarization parameters
$S$ and $T$. With a lower value for the top quark mass this is perhaps
becoming questionable, but, more importantly, there are many other very
simple models that also easily account for precision electroweak data
and have cutoff scales significantly larger than that of the Standard
Model.  It is not clear that the Standard Model, with the lowest
cutoff scale, is to be preferred.  This is especially true since these
other models can contain stable Weakly Interacting Massive Particles
(WIMPs) that naturally account for the amount of dark matter (DM) observed
in the universe.  If one of these alternative electroweak models
describes nature, and has an energy cutoff above say 2 TeV, the
consequences for the LHC are enormous. The LHC may not be able to
probe the new ``canceling physics'' at the cutoff, but may instead
discover and explore the new electroweak model that has Improved
Naturalness compared to the Standard Model~\cite{Barbieri:2006dq}.

The simplest way to construct an electroweak sector with Improved
Naturalness is to make the Higgs boson heavy, for example in the range
of 400 to 600 GeV, since the
expectation for the cutoff scale from top quark loops increases
linearly as $ \sim 3.7\,m_H$~\cite{Barbieri:2005ri}. The model must also include some extra
states that give a contribution to the $T$ parameter in the
right range to account for the experimental value, given that the
Higgs is heavy~\cite{Chivukula:2000px}. This is typically accomplished if the new states have
masses of a few hundred GeV and dimensionless couplings of order
unity, and does not require any precise parameter tunings. Such states
are well known to be ideal WIMP candidates.

If the LHC discovers a heavy Higgs, what other signals will point to a
model of Improved Naturalness?  It could be that several of the new
states are colored, as for example in the case of ``$\lambda$SUSY''~\cite{Barbieri:2006bg},
and will be copiously produced at the LHC. However,
unlike the case of the Standard Model
with a light Higgs and a low cutoff, there is no expectation that new
colored states will be accessible to LHC.  Neither a contribution to
$T$ nor WIMP dark matter argue that the new states should be
colored; one simple model involves only an extra ``Inert'' Higgs
doublet~\cite{Barbieri:2006dq}.  It is quite likely that the new states have only electroweak
interactions. 

In this paper we introduce and study perhaps the simplest model of
Improved Naturalness. There is just a single Higgs doublet so that
the naturalness analysis is essentially the same as
the familiar Standard Model case.  The simplest possibility for the
new states appears to be a single heavy vector lepton doublet
$(L,L^c)$, but this does not contribute to $T$ and is excluded as dark
matter. Hence in addition we add a heavy neutral Majorana state, $N$, 
that mixes with the vector lepton doublet via Higgs couplings. These
couplings introduce a contribution to $T$ and allow the lightest
neutral mass eigenstate to be the dark matter.  Such a model can have
a natural energy cutoff as high as 2 TeV, and provides a very clean
environment for studying signals for the new electroweak states
at the LHC, and also at direct dark matter detectors.

%============================================================================== 
\section{A model with heavy leptons}\label{Section:Model}
%============================================================================== 

An increase in the Higgs mass introduces a large negative contribution to 
$\Delta T$, so that the Standard Model must be extended to compensate for 
this change. The minimal extension that also provides a dark matter candidate, 
introduces one vector-like fermion doublet pair and one fermion singlet\footnote{For convenience we call these fermions ``leptons'', as they are charged under $\su(2)$ and do not carry color. They do not, however, carry lepton number and do not mix with or interact with the SM leptons. They could equally well be called, \eg,  higgsinos and bino.}. The singlet is essential for two reasons:\ pure doublet dark matter is already ruled out by direct detection \cite{Cirelli:2005uq}, and the isodoublets must be split in order to contribute to $\Delta T$. These extra states are odd under a discrete $Z_2$ symmetry with the SM states being even; this forbids any mixing with the SM leptons and makes the lightest new state stable and a possible dark matter candidate.

The new interactions, allowed by gauge and discrete symmetries, are
\be\label{eqn:lagrangian}
\Delta \mathcal{L}= -\lambda L H N -\lambda' L^c \tilde{H} N + M_L L L^c + \frac{1}{2}M_N N^2 + h.c.\,,
\ee
where 
\be
L=\dbinom{\nu}{E}, \quad L^c=\dbinom{E^c}{\nu^c},
\ee
$N$ is a SM singlet, and $\tilde{H}=i\sigma_2 H^*$.  Once electroweak symmetry is broken the Yukawa interactions contribute to the masses of the neutral states, splitting the $\su(2)$ doublets.  The mass matrix after electroweak symmetry breaking, in the basis $(N,\nu, \nu^c)$, is
\be
\begin{pmatrix}
M_N & \lambda v & \lambda' v \cr
\lambda v & 0 & M_L \cr
\lambda'v & M_L & 0
\end{pmatrix},
\ee
where our normalization is such that the Higgs vev is $v=174\,\gev$.  In general, the new parameters can be complex but one can always perform field redefinitions such that the only physical phase appears in the Majorana mass term.  For simplicity we take this term to be real.
(Models with the same field content have been studied in relation to dark matter and unification~\cite{Mahbubani:2005pt,D'Eramo:2007ga}, but with different context and motivation.)

Upon diagonalization of the above mass matrix, the neutral fermions $N,\nu, \nu^c$ yield three Majorana ``neutrinos'' $(\nu_1,\nu_2,\nu_3)$, labeled by ascending mass. Furthermore the  charged  fermions $E,E^c$ will form a heavy charged ``lepton'' $\psi_E$ of mass $M_L$.  Whenever $m_{\nu_1}<m_{E}$ the lightest parity-odd state is $\nu_1$, which is a good DM candidate because of the $Z_{2}$ symmetry.  The mass eigenstates are related to the interaction states by 
\be
\begin{pmatrix}
\nu_1 \cr
\nu_2 \cr
\nu_3
\end{pmatrix}
=
\begin{pmatrix}
\vartheta_1 & \alpha_1 & \beta_1 \cr
\vartheta_2 & \alpha_2 & \beta_2 \cr
\vartheta_3 & \alpha_3 & \beta_3 
\end{pmatrix}
\begin{pmatrix}
N \cr
\nu \cr
\nu^c
\end{pmatrix}.
\ee
In terms of mass eigenstates the interactions are
\begin{align}
\mathcal{L_\text{int}} = &     
\frac{g}{2 c_W}  
\bar \psi_{\nu_i} \gm \gf V_{ij} \psi_{\nu_j} \,  
Z_\mu 
+
\frac{g}{c_W} \bar \psi_{E} \gm \psi_{E}  (T^3-s_W^2 Q ) \, Z_\mu 
+ e Q \, \bar \psi_{E} \gm \psi_{E} A_\mu \nn \\  
+&\frac{g}{\sqrt{2}} 
 \bar \psi_{\nu_i} \gm \left(g_V^i-g_A^i\gf\right) \psi_{E} \, W^+_\mu 
+\frac{g}{\sqrt{2}} 
 \bar \psi_{E} \gm \left(g_V^i-g_A^i\gf\right) \psi_{\nu_i} \, W^-_\mu 
\nn \\  
+& U_{ij} \,\bar \psi_{\nu_i} \psi_{\nu_j} h,  
\label{Lint2}
\end{align}
where
\begin{align}
V_{ij} &= \alpha_i \alpha_j - \beta_i \beta_j \label{Coupling_V}\\
U_{ij} &= \frac{1}{\sqrt 2}(\lambda \alpha_i  +\lambda' \beta_i) \vartheta_j + (i \leftrightarrow j)  \label{Coupling_U} \\
g_V^i &= \frac{\alpha_i+\beta_i}{2}  \label{Coupling_gV}\\
g_A^i &= \frac{\alpha_i-\beta_i}{2}  \label{Coupling_gA}.
\end{align}
The expressions for the masses and mixing angles in terms of the four 
parameters of (\ref{eqn:lagrangian}) are given in  Appendix A.

\subsection{Naturalness and perturbativity}

Since our model has a single Higgs doublet, it has naturalness
properties that are very similar to the Standard Model.
The quadratically divergent one-loop corrections to the Higgs
mass parameter are no larger than the tree-level contribution provided
they are cutoff by~\cite{Barbieri:2006dq}
\begin{align}
&\Lambda_{H} \sim 1.3 \text{ TeV}  \\
&\Lambda_{t} \sim 3.7 m_{H} \sim 1.85 \text{ TeV}  \\
&\Lambda_{g} \sim 8.7 m_{H} \sim 4.35 \text{ TeV} 
\end{align}
for Higgs, top, and gauge boson loops, respectively. The top and gauge contributions
are evaluated for a 500 GeV Higgs. The new Yukawa
couplings also contribute a quadratic divergence, leading to a
naturalness cutoff
\be
\Lambda_{\lambda} \simeq \frac{\sqrt{3}}{\lambda}
\Lambda_t,
\ee 
the same expression is true for $\lambda \leftrightarrow \lambda'$, so that we consider $\lambda,\lambda' <3$.

We aim to have a completely natural and calculable theory up to $\sim 2$ TeV,
meaning that there are no new \emph{colored} states accessible to LHC\footnote{The cutoff for the Higgs loops is below 2 TeV meaning that the Higgs sector will also be extended in a completely natural theory, or there is a small amount of tuning in the Higgs sector.  However, such states need only couple to the Higgs and will not be a seen at the LHC.}. This requires that our model remain perturbative up to 2 TeV.  Note that if one assumes that all the cutoffs take a common value \cite{Veltman:1980mj} then the quadratic divergence in the Higgs mass cancel in the SM, for $m_H\sim 300\,\gev$.  Without knowing the particular UV completion of the SM that cancels the divergences there is no reason\cite{Barbieri:2006dq} \emph{a priori} to assume that these scales are all the same, the canceling physics for each sector can enter at different scales.

In addition to 
implying higher naturalness cutoff scales, a heavy Higgs also implies a larger 
quartic self-coupling. 
To determine the behavior of the Higgs quartic and other couplings, we first run the 
Standard Model gauge couplings from their values at the weak scale 
up to a few hundred GeV, at which point the effects of the heavy 
leptons become important. Specifically, we choose a ``matching scale'' 
of $\mu_{\rm m}=1.36 m_{H}$, at which we impose the condition $\lambda_{H}(\mu_{\rm m}) 
= m_{H}^{2} / 4 v^{2}$, since this relation includes 1-loop threshold 
corrections \cite{Barbieri:2006dq}. In addition, for simplicity we impose 
the following boundary conditions: $\lambda_{t}(\mu_{\rm m}) = 171.4\,\gev / v$ 
for the top Yukawa coupling; the gauge couplings $g_{i}(\mu_{\rm m})$ 
equal the results obtained from running the Standard Model RGE's~\cite{Machacek:1981ic} up 
to $\mu_{\rm m}$; and the heavy lepton Yukawa couplings $\lambda(\mu_{\rm m})$ 
and $\lambda'(\mu_{\rm m})$ are chosen by hand. We then run the 
one-loop RGE's for these couplings, taking into account the effects 
of the heavy leptons.

\begin{figure}[t]
\begin{center}
\epsfig{file=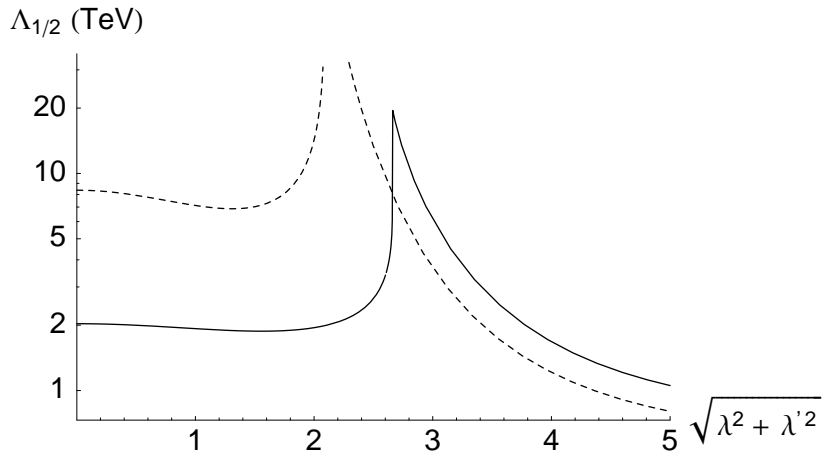, width=0.7\columnwidth}
\caption{The energy scale $\Lambda_{1/2}$, in TeV, indicating the 
onset of non-perturbativity for the Higgs quartic, as a function 
of $\sqrt{\lambda^{2}(\mu_{\rm m}) 
+ \lambda'^{2}(\mu_{\rm m})}$. The solid curve is for a $500$ 
GeV Higgs, and the dashed curve for $400$ GeV.
}\label{Hcut}
\end{center}
\end{figure}

We estimate the onset of the breakdown of perturbation theory 
by determining when the Higgs quartic RGE becomes ${\cal O}(1)$. Since the running is dominated by the Higgs quartic self-coupling we define a breakdown scale $\Lambda_{1/2}$ as
\be
\left| \frac{\partial \ln \lambda_{H}(\mu)}{\partial \ln \mu}\right|_{\Lambda_{1/2}}\sim \frac{3}{2\pi^2}\lambda_H(\Lambda_{1/2})=\frac{1}{2},
\ee
where $\mu$ is the renormalization scale.
This scale depends on the value of the Higgs mass (\ie, 
the initial value chosen for the Higgs quartic) as well as the 
initial values chosen for the lepton Yukawa couplings $\lambda$ 
and $\lambda'$. Fig.~\ref{Hcut} shows  
$\Lambda_{1/2}$ as a function of initial Yukawa couplings 
(in the form $x \equiv \sqrt{\lambda^{2} + \lambda'^{2}}$, 
as they appear in the RGE's). For intermediate values of $x$, 
cancellations occur in the RGE for $\lambda_{H}$, between Standard 
Model and new physics contributions. These cancellations allow 
$\lambda_{H}$ to remain perturbative to much higher energies, 
as indicated by the peak in the plot. We performed a similar 
analysis for the heavy lepton Yukawa couplings, $\lambda$ and 
$\lambda'$. The Higgs quartic reaches the limit of perturbativity, 
however, well before the Yukawa couplings. It is sufficient, 
therefore, only to consider the Higgs quartic here.

In this paper we consider values for $\lambda$ and $\lambda'$ 
that give $x<3$ (see, \eg, Table~\ref{tab:points} below), so that 
for a $500$ GeV Higgs the theory remains 
perturbative up to at least $2$ TeV. 
However Fig.~\ref{Hcut} suggests two different schemes for a UV completion: 
if it occurs at 2 TeV, the effective theory below 2 TeV is
completely natural, but if the physics at 2 TeV is strongly
interacting then its direct contributions to the electroweak precision tests must be about an order of magnitude smaller than naive
estimates.  Alternatively, if the new Yukawa couplings give $x$ near 3, 
our model could describe physics up to 5 TeV.  In this case the
Little Hierarchy problem is solved, but about an order of magnitude
fine tuning is required in the overall expression for the Higgs mass
parameter. Either possibility has an order of magnitude less fine tuning than
the Standard Model with a light Higgs, but they have very different implications for the LHC, as will be discussed later.

%============================================================================== 
\section{Experimental bounds on the model}\label{Section:Bounds}
%============================================================================== 

The model has only four free parameters in addition to the Higgs mass, which we take to be around $500\,\gev$.  Requiring that the model satisfies experimental bounds from electroweak precision tests (EWPT) and that $\nu_1$ has escaped direct detection and provides the correct dark matter abundance places constraints on the parameter space of the model.  We discuss each constraint in turn and show that there is a large region of parameter space which is allowed.  We will describe the general features of the model in this region.

\subsection{Electroweak precision tests}

\begin{figure}[t]
\begin{center}
\begin{picture}(475,100)(25,0)
\Photon(75,50)(125,50){4}{5.5}
\Photon(175,50)(225,50){4}{5.5}
\ArrowArc(150,50)(25,0,180)
\ArrowArc(150,50)(25,180,360)
\Text(150,15)[]{$\nu_i$}
\Text(150,85)[]{$\nu_j$}
\Text(65,50)[]{$Z$}
\Text(235,50)[]{$Z$}
\Photon(275,50)(325,50){4}{5.5}
\Photon(375,50)(425,50){4}{5.5}
\ArrowArc(350,50)(25,0,180)
\ArrowArc(350,50)(25,180,360)
\Text(350,15)[]{$E^\pm$}
\Text(350,85)[]{$\nu_i$}
\Text(263,50)[]{$W^\pm$}
\Text(439,50)[]{$W^\pm$}
\end{picture}
\end{center}
\caption{Contributions to $\Delta T$ from the heavy lepton model.}\label{fig_deltat}
\end{figure}
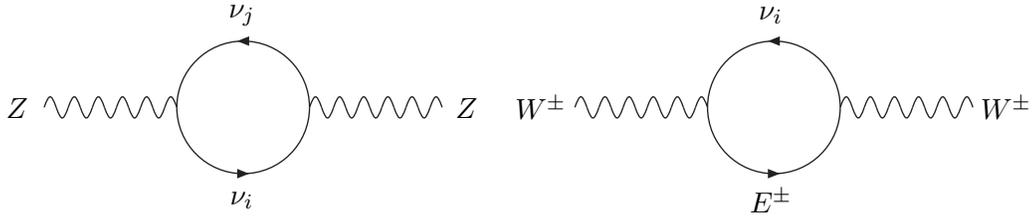

The oblique corrections to the gauge boson propagators coming from integrating out heavy states are well described by the $S$, $T$ and $U$ parameters  \cite{Peskin:1991sw}.  A 300--600~GeV Higgs leads to corrections to the $S$ and $T$ parameters of the form,
\begin{align}
\Delta T \approx -\frac{3}{8\pi\cos^2\theta_W}\log\frac{m_h}{m_Z}\approx -0.25\pm 0.05
&&
\Delta S \approx \frac{1}{6\pi}\log\frac{m_h}{m_Z}\approx 0.08\pm 0.02.
\end{align}
The large correction to $T$, in particular, must be canceled by the new leptons.  The contributions to the vector boson self-energies  from the additional leptons  are shown  in Fig.\ \ref{fig_deltat}. The contributions to the $S$ and $U$ parameters are much smaller than those to the $T$ parameter and will be neglected.

\begin{figure}[th]
\begin{center}
\epsfig{file=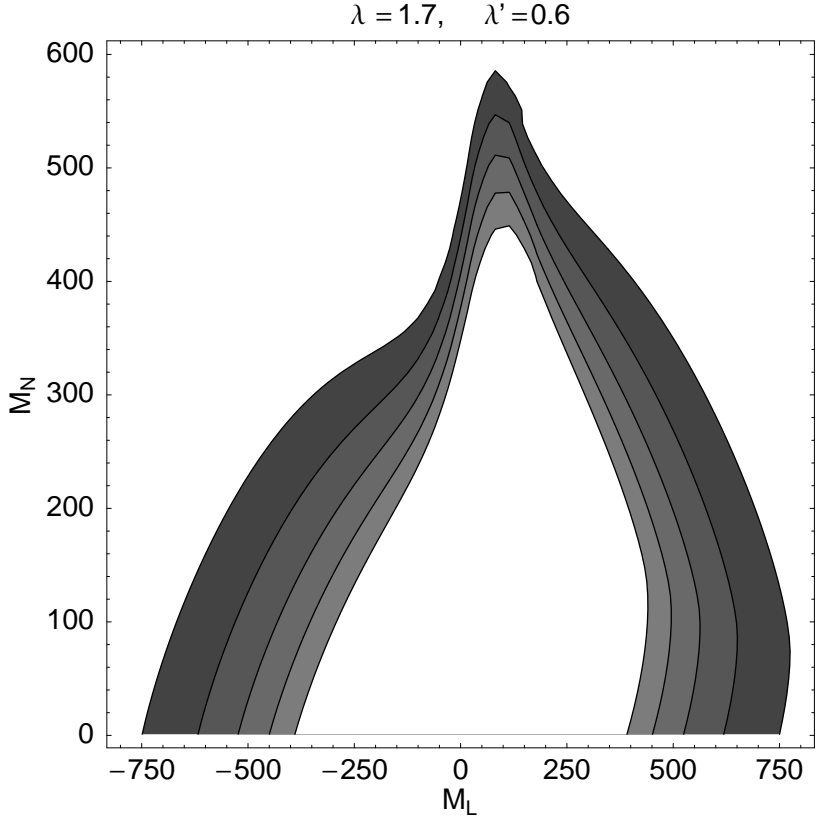, width=0.7\columnwidth}
\caption{Contour plot showing the allowed region for the doublet mass $M_L$ and singlet mass $M_N$ for one choice of Yukawa couplings, $\lambda=1.7$ and $\lambda'=0.6$. The gray regions allow a heavy Higgs, \ie, they have a contribution to $\Delta T$ in the range $[0.15,0.35]$. The outer boundary corresponds to $\Delta T=0.15$ and the inner boundary to $\Delta T=0.35$, with the contour lines in the gray region equally spaced. 
}\label{Tplot}
\end{center}
\end{figure}

The change $\Delta T$ in the $T$ parameter is defined as \cite{Peskin:1991sw}
\be
\Delta T = \frac{\Pi_{WW}(0) - c_W^2 \Pi_{ZZ}(0)}{\alpha_{\text{em}} m_W^2}. 
\ee
where the contributions to the vacuum polarizations from a fermionic $SU(2)$ doublet is given in \eg~\ Ref.~\cite{Lavoura:1992np}. The exact expression for $\Delta T$ in our model is not very illuminating, as it depends on many mixing angles.  However, the contribution of the new leptons disappears when $\lambda=\lambda'$ since in this limit one can extend the custodial $SU(2)$ symmetry protecting $T$ to the new leptonic sector.  Thus, parametrically $\Delta T\sim \left( \lambda^2-\lambda'^2\right)^2$, and we find that in order to get a sufficiently large contribution $ \left( \lambda^2-\lambda'^2\right)\gtap 1$.  As an example in Fig.~\ref{Tplot} we plot the region allowed by $\Delta T \in [0.15,0.35]$ in the $(M_{L},M_{N})$ plane for a particular choice of the Yukawa couplings.

\subsection{Dark matter abundance}

Parametrically the annihilation cross section of the DM scales in a similar fashion to $\Delta T$ since the cross section behaves as $\sigma\sim V_{11}^2$, with (for $M_L\gg M_N$) $V_{11}\sim \left(\lambda^2-\lambda'^2\right) \frac{v^2}{M_L^2}$.  Given the lower bound, discussed in the previous section, on the difference of the Yukawa couplings $\lambda$ and $\lambda'$ due to the EWPT constraint, there is a lower bound on $M_L$ since low values lead to too little dark matter.  We find this lower bound to be around $250\,\gev$.

The lightest of the three neutrino states is stable and is our WIMP dark matter candidate, which we denote by $\chi$ ($=\nu_1)$. The computation of the abundance requires evaluating the annihilation cross sections for all three allowed final states, $f\bar f$, $W^+ W^-$ and $Z Z$, where $f$ is any SM fermion.  Because of its large mass, contributions coming from exchanging the Higgs will be negligible.

To obtain the dark matter abundance we compute the thermally averaged annihilation cross sections for the various annihilation channels, as well as the freeze-out temperature at which the WIMPs decoupled from thermal equilibrium.
As usual the standard approximate solution of the Boltzmann equation \cite{Kolb:1990vq,Bertone:2004pz} leads to
\be
\Omega_\chi h^2 \approx
\frac{1.07 \times 10^9 \text{GeV}^{-1}}{M_{\text{Pl}}} \,
\frac{x_F}{\sqrt{g_*}} \,
\frac{1}{a+3b/x_F}
\ee
where $a$ and $b$ are obtained by expanding the thermally averaged annihilation cross section in powers of the relative velocity of the DM particles, and are given in Appendix B for the various annihilation channels.
The freeze-out temperature $T_F=m_\chi/x_F$ is obtained by solving
\beq
x_F = \ln \left[c(c+2) \sqrt{\frac {45}{ 8}} \,
\frac{g}{2\pi^3} \,
\frac{m_\chi M_{\text{Pl}} (a+6b/x_F)}{\sqrt{g_* x_F}} \right],
\eeq
where $c$ is some constant of order one, and is typically near
$m_\chi/25$.

\begin{figure}[t]
\begin{center}
\epsfig{file=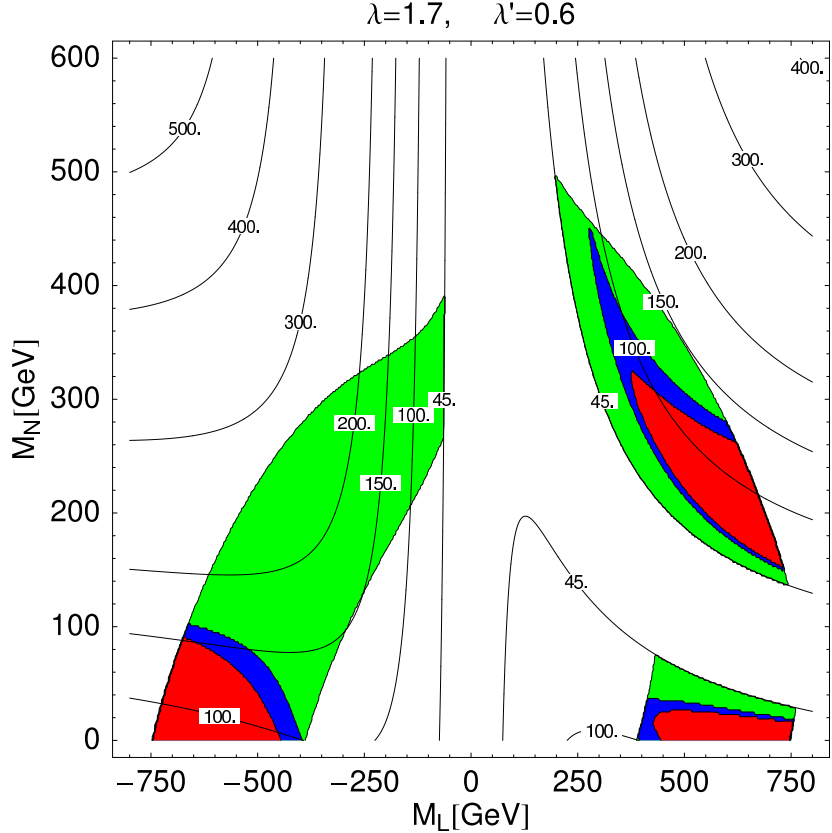, width=0.7\columnwidth}
\caption{Parameter regions as a function of the doublet mass $M_L$ and the singlet mass $M_N$ for one choice of Yukawa couplings, $\lambda=1.7$ and $\lambda'=0.6$. The predicted dark matter abundance agrees with the observed value in the dark blue bands, but is too large (small) in the red (green) regions. The shaded regions lead to a heavy Higgs boson, as can be seen by comparing with Fig.~\protect\ref{Tplot}. The labeled contours give the dark matter particle mass in GeV.}\label{DMplot}
\end{center}
\end{figure}

As we will show in the following, in our case there is always sufficient separation between the mass of the DM particle and the next heaviest state so that one can safely neglect coannihilation contributions \cite{Griest:1990kh}.

In addition to the analytic calculations, we have used the program micrOMEGAs~\cite{Belanger:2006is} (which includes all possible tree level diagrams) to numerically evaluate the DM abundance, finding good agreement between these two calculations.  In our scan of parameter space (see below) we use micrOMEGAs to compute the abundance.  In Fig.~\ref{DMplot} we show an example of the region allowed by dark matter, 
and by EWPT with a heavy Higgs, for the same choice of Yukawa couplings as in Fig.~\ref{Tplot}.

\subsection{General properties of parameter space}

We have numerically sampled the region of parameter space that is allowed by EWPT and dark matter, collecting $\mathcal{O}(10^4)$ allowed points. To be specific, we require a contribution to $\Delta T$ in the range 0.15--0.35 and a dark matter abundance $\Omega h^2$ in the WMAP range 0.09--0.13~\cite{Spergel:2006hy}. A lower value would of course be allowed, but it would require another component of dark matter, such as axions. We would like to consider the case where our model is the only new physics needed.  Finally, we also require that the mass of the lightest neutrino is larger than 45 GeV to escape detection at LEP.

\begin{figure}[t]
\begin{center}
\epsfig{file=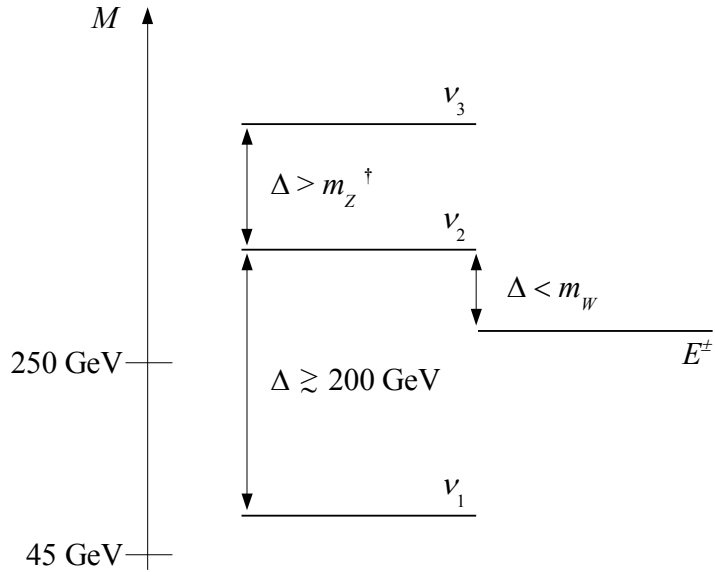, width=0.6\columnwidth}
\caption{General features of the lepton spectrum. The $^\dagger$ means that this inequality is only true over approximately half of parameter space. \label{fig:spectrum}}
\end{center}
\end{figure}

From this analysis we find some general properties of the particle spectrum. The doublet mass $M_L$ (\ie, the charged lepton mass) is always larger than 250--300 GeV, because of the requirement of dark matter, but there is no lower bound on the singlet mass $M_N$. As pointed out in Sect. 3.1 in order to obtain a large enough $\Delta T$ the two Yukawa couplings must not be too close,  since they are constrained by $ \left( \lambda^2-\lambda'^2\right)\gtap 1$.
Moreover $\nu_2$ is always heavier than the charged lepton but never by more than $m_W$, so there will be no decays $\nu_2 \to E W$.
The mass splitting between $\nu_3$ and $\nu_2$ is larger than $m_Z$ in about half of parameter space, allowing the decay $\nu_3 \to \nu_2 Z$.  Similarly, the mass difference between $\nu_3$ and $E$ is almost always larger than $m_W$, so that the decay channel $\nu_3 \to E W$ is open.  Finally, the mass splitting between the two lightest neutrinos is almost always larger than 200 GeV and is larger than 300 GeV in more than 90\% of parameter space.  These features of the spectrum are illustrated in Fig.~\ref{fig:spectrum}.

\section{Direct detection of dark matter}\label{sec:DDtheory}

In addition to cosmological bounds on dark matter abundance, there may be constraints on its properties coming from direct detection searches of WIMPs in the halo of the Milky Way.  By looking for WIMP--nucleus scattering as the Earth moves through the WIMP background several experiments have obtained upper bounds on scattering cross sections, as a function of WIMP mass.  There are two contributions to the cross section, spin-independent and spin-dependent scattering. At present the experimental bounds are stronger for the spin-independent cross section. Due to the smallness of the coupling of the lightest neutrino to $Z$ and the large mass of the Higgs we do not expect the present bounds to constrain our model.  However, for some regions of parameter space we may expect a discovery in the 2007 CDMS run, while the most sensitive experiments planned for the future would be able to discover or to rule out most of the parameter space.

Using the the method presented in \cite{Tovey:2000mm} 
(see also \cite{Jungman:1995df,Ellis:2000ds,Ellis:2001pa}) to compare results from different experiments, we calculate the ``standard'' WIMP scattering cross section, $\sigma_0$, on nucleons and compare it to the experimental bounds. This standard cross section (actually the WIMP-nucleon cross section at zero momentum transfer) is defined by
\be
 \frac{d\sigma(q=0)}{d|\vec q|^2} \equiv \frac{\sigma_0}{ 4 \mu^2 v^2},
\ee
where $\mu= m_N m_\chi/(m_N + m_\chi)$ is the reduced mass of the nucleon-WIMP system ($m_N$ is the nucleon mass), $v$ is the WIMP velocity in the lab frame, and $q$ is the  momentum transfer.

Integrating out the vector bosons and the Higgs in (\ref{Lint2}), one obtains an effective Lagrangian for neutrino--quark scattering
\be 
{\cal L} = 
b_q^V \, \bar\chi \gm \gf \chi \; \bar q \gamma_\mu q + 
b_q^A \, \bar\chi \gm \gf \chi \; \bar q \gamma_\mu \gf q + 
f_q \, \bar \chi \chi \, \bar q q,
\label{effchiq}
\ee
where $\chi$ represents the dark matter particle and $q$ any quark and the various couplings are given by
\begin{align}
b_q^V&=  C_V V_{11}\dfrac{e^2}{4s_W^2c_W^2m_Z^2}\label{DDbqV}\\
b_q^A&=- C_A V_{11}\dfrac{e^2}{4s_W^2c_W^2m_Z^2}\label{DDbqA}\\
f_q &=  U_{11} \frac{e \, m_q }{2 s_W m_W m_H^2}\label{DDfq}
,
\end{align}
with $C_V=T_3-2Q s_W^2$ and $C_A=T_3$.
Higgs exchange leads to a spin-independent cross section, whereas $Z$ exchange leads to both a spin-dependent cross section through the coupling to the axial-vector quark current, and to a spin-independent cross section through the quark vector current. The latter spin-independent contribution is usually neglected in standard MSSM studies because it is proportional to the WIMP velocity squared. One may worry that it is not negligible in our case because of the much heavier Higgs, but we will show below that it can be safely neglected in our calculation too.

\subsubsection*{Spin-independent interaction (SI)}

The spin-independent interaction is mediated by both Higgs and $Z$ exchange. 
The contribution to the scattering amplitude from Higgs exchange  comes from the last term of \eq{effchiq}.
Neglecting ${\cal O}(v^2)$ terms, the ``standard'' WIMP-nucleon SI cross section is
\be
\sigma_0^N = \frac{4}{\pi} \mu_N^2 f_N^2,
\ee
where $\mu_N$ is the reduced mass and $f_{N}$ can be expressed in terms of the matrix elements of the operators $\bar q q$ and $G_{\mu\nu}G^{\mu\nu}$ between nucleon states. These matrix elements can be derived using QCD sum rules~\cite{Jungman:1995df,Ellis:2000ds,Ellis:2001pa}:\ recent numerical values can be found in \cite{Ellis:2000ds,Ellis:2001pa}. Using these numbers we find
\be
f_N\approx 0.3 \, U_{11}\frac{g \, m_N}{2m_W m_H^2}.
\ee
To very good approximation the Higgs has the same coupling to protons and neutrons. We take them to be equal and consider only one cross section $\sigma^p$,  
\be
\sigma^p  \simeq 9.37 \times 10^{-45} \, U_{11}^2 
\left( \frac{500 \text{ GeV}}{m_H}\right)^4\text{  cm}^2, 
\label{eq:SIformula}
\ee
independent of the WIMP mass to within a few percent.

The $Z$ exchange contribution to the SI cross section is given by the first term in \eq{effchiq}.
Because the $Z$ couples coherently to the quarks one can replace the quark couplings with the corresponding nucleon couplings. If we form the ratio 
of the squared amplitudes for $Z$ and Higgs exchange, it turns out to be on the level of a few percent times the ratio of couplings $(V_{11}/U_{11})^2$. Now, $(V_{11}/U_{11})^2$ is always smaller than one, and usually much smaller, so we therefore conclude that the $Z$-exchange contribution to the SI cross section can be neglected.

\subsubsection*{Spin-dependent interaction (SD)}

To compute the spin-dependent scattering on a nucleon $N=p,n$ we need the matrix element
\be
\bra{N} \bar q \gamma_\mu \gf q \ket{N}  = 2s_\mu^{(N)} \Delta q^{(N)}
\label{SDME}
\ee
where $s_\mu^{(N)}$ is the spin vector of the nucleon and $\Delta q^{(N)}$ specify the amount of spin carried by the quark $q$ inside the nucleon $N$; they are combinations of the first moment of the polarized structure function $g_1(x,Q^2)$ and are known from experiment (see \eg~\cite{Mallot:1999qb}).

From Eqs.\ \eq{effchiq}, \eq{DDbqA} and \eq{SDME} the effective Lagrangian for the WIMP coupling to nucleons is 
\be 
{\cal L} = \dfrac{e^2 V_{11}}{4s_W^2c_W^2m_Z^2} \;
\bar\chi \gm \gf \chi \; \bar N s_\mu^{(N)}  N
\sum_{q=u,d,s} 2 T^{3(q)} \Delta q^{(N)},
\ee
where $T^{3(q)}$ is the isospin of the quark. Using this interaction and the Wigner--Eckart theorem leads to the standard cross section
\be
\sigma_0^N = \frac{24G_F^2}{\pi} 
\mu_N^2 a_N^2
\ee
where, using values for the parameters $\Delta q^{(p,n)}$ to be found in \eg~\cite{Mallot:1999qb,Ellis:2000ds,Ellis:2001pa},
\bea
a_p&=&0.705 \, V_{11} \\
a_n&=&-0.555 \, V_{11}.
\eea
Thus, we find that
\begin{align}
\sigma^p & \simeq 1.77 \times 10^{-37} \, V_{11}^2 \text{  cm}^2 
\label{eq:SDformula1}\\
\sigma^n & \simeq 1.10 \times 10^{-37} \, V_{11}^2 \text{  cm}^2. 
\label{eq:SDformula2}
\end{align}

\begin{figure}[t]
\begin{center}
\epsfig{file=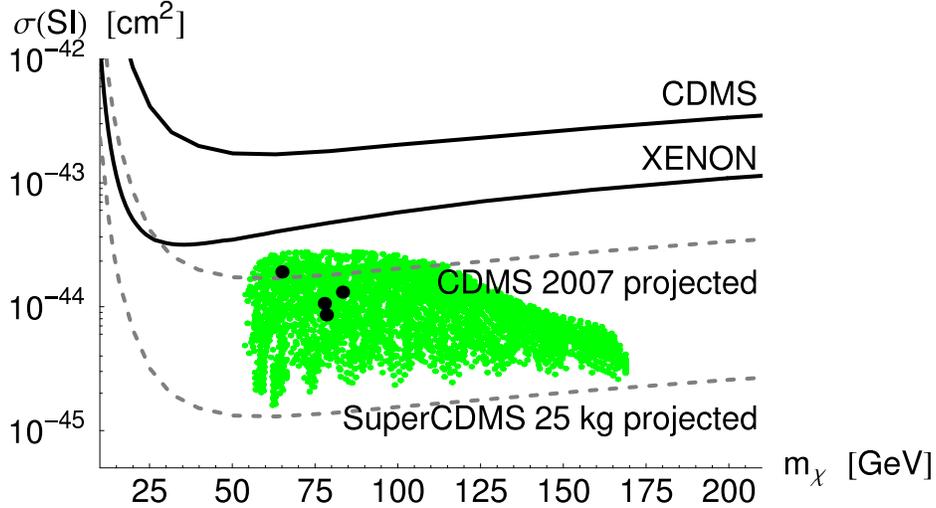, width=0.8\columnwidth}
\caption{Cross sections for spin-independent dark matter searches. The green region is our parameter space for $m_H = 500$ GeV. For a Higgs mass of 400 GeV, the predicted cross section is increased by a factor of about 2.5. The upper solid black curve is the present CDMS exclusion limit~\protect\cite{Akerib:2005kh} and the lower solid curve is the XENON limit~\protect\cite{Angle:2007uj}. The upper dashed curve is the projected CDMS II  limit from the 2007 run. The lower dashed curve is the projected bound from the first 25 kg-phase of the planned SuperCDMS experiment. 
The black points represent our LHC phenomenology points from Table~\protect\ref{tab:points}.}\label{DD_SI_plot}
\end{center}
\end{figure}

\subsection*{Experimental searches}

We now use the results (\ref{eq:SIformula}, \ref{eq:SDformula1}, \ref{eq:SDformula2}) of the previous two subsections  to see what are the current constraints on our model coming from the experimental searches.  The two strongest existing bounds on spin-independent WIMP--nucleon scattering are currently from the XENON~\cite{Angle:2007uj} and CDMS~\cite{Akerib:2005kh} experiments. 
For spin-dependent scattering the bounds are somewhat weaker; the strongest bound on WIMP--neutron scattering is from CDMS~\cite{Akerib:2005za}, while the strongest bounds on WIMP--proton scattering\footnote{One can derive an indirect bound from SuperK data by searching for neutrinos coming from come from annihilation of gravitationally trapped WIMPs. This bound has been derived in \cite{Desai:2004pq} in the case of SUSY models, and we leave for future study the same analysis for our model.} come from two experiments:\ the NAIAD experiment~\cite{Spooner:2000kt,Alner:2005kt} and Shimizu \emph{et al.}~\cite{Shimizu:2005kf}. These two last bounds are essentially the same, so we will for simplicity only consider the NAIAD results.

\begin{figure}[t]
\begin{center}
\epsfig{file=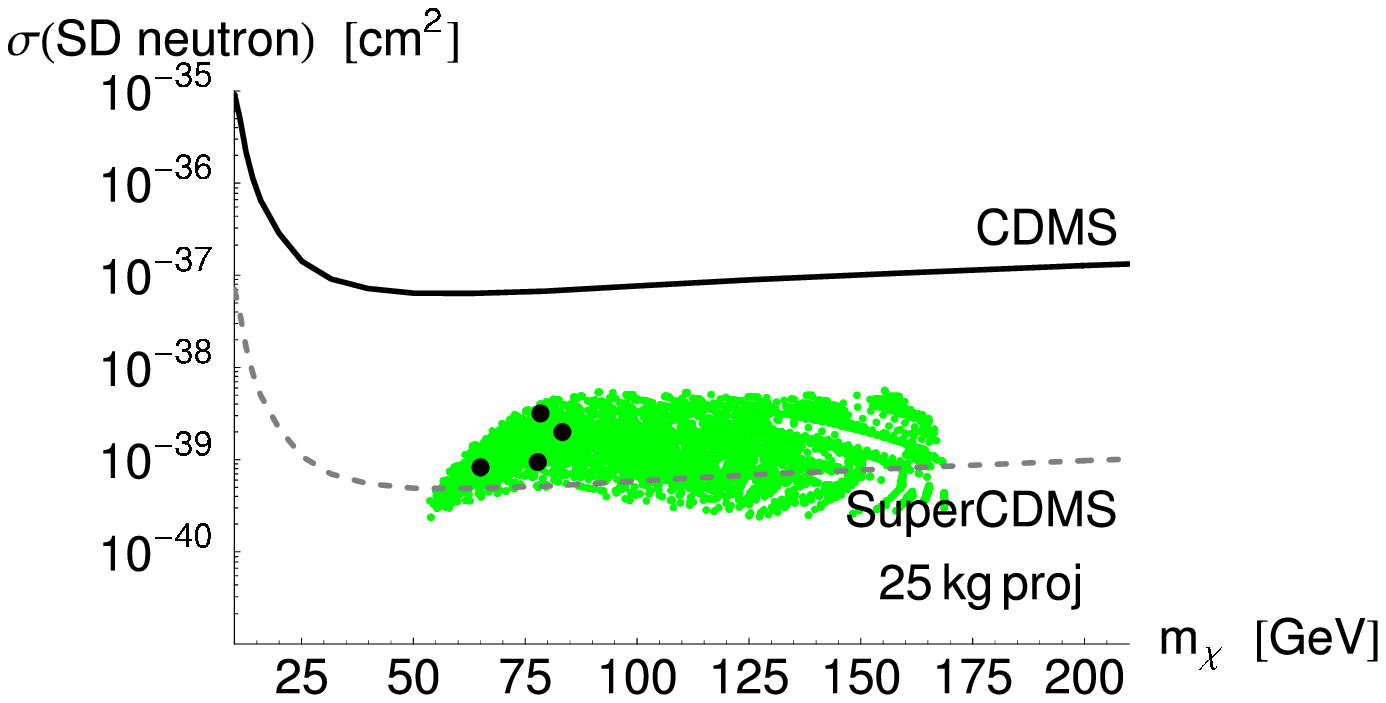, width=0.8\columnwidth}
\epsfig{file=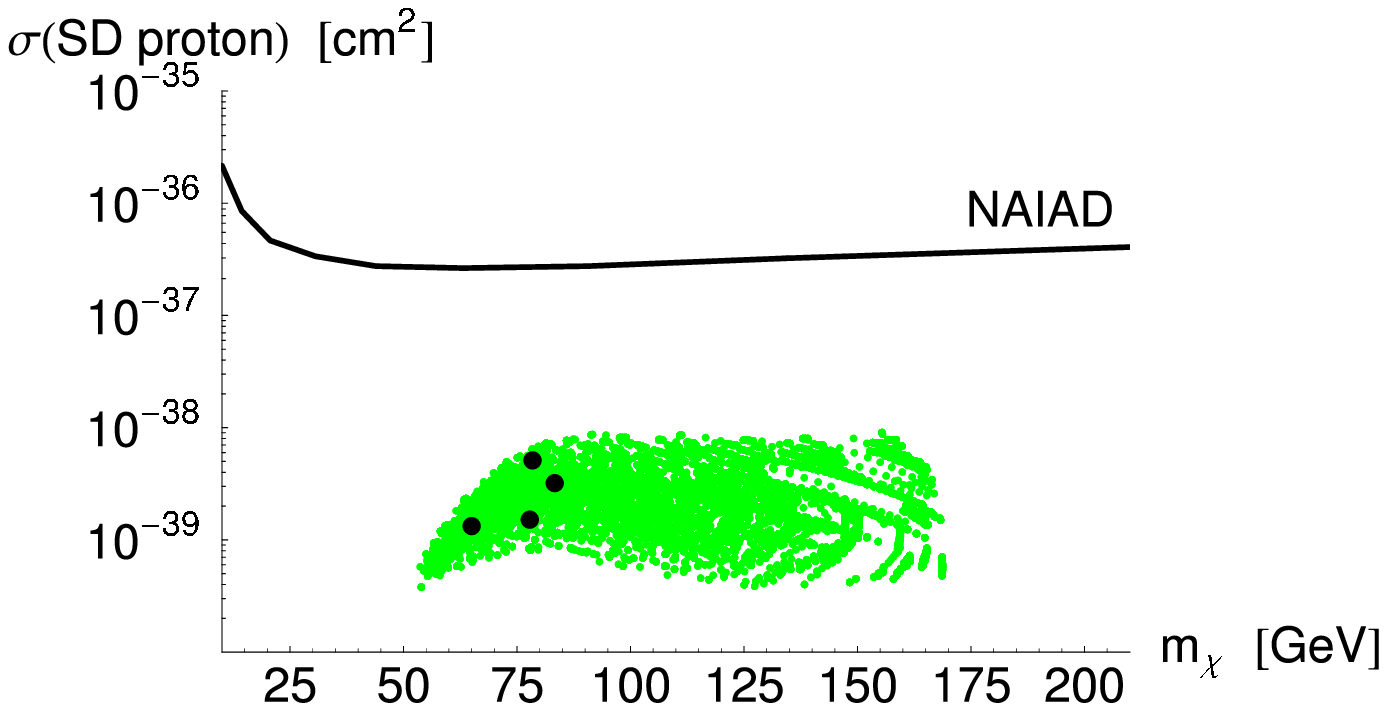, width=0.8\columnwidth}
\caption{Cross sections for spin-dependent dark matter searches for WIMP--neutron scattering (upper) and WIMP--proton scattering (lower). The green regions are our parameter space. The solid black curves are the present CDMS~\protect\cite{Akerib:2005za} and NAIAD exclusion limits. The dashed gray curve in the neutron plot is the projected SuperCDMS limit. }\label{DD_SD_plot}
\end{center}
\end{figure}

In the near future, the 2007 CDMS run is projected to increase the sensitivity of spin-independent scattering by an order of magnitude. Further on, the EURECA, SuperCDMS, XENON, and ZEPLIN experiments all expect to measure cross sections down to $10^{-46}$ cm$^2$ per nucleon with ton-scale detectors \cite{Akerib:2006ks}. The first proposed phase of SuperCDMS plans to use 25 kg of detector material and start operating in 2011~\cite{Akerib:2006rr}.

Using the results derived above we have performed a scan over the allowed region of parameter space numerically computing the standard cross sections $\sigma_0$ for all points in parameter space that are allowed by EWPT and dark matter abundance, as described in Sect.~\ref{Section:Bounds}. The results are shown as a function of the WIMP mass in Fig.~\ref{DD_SI_plot} for SI scattering and in Fig.~\ref{DD_SD_plot} for SD scattering together with corresponding experimental search limits and projected limits. The experimental limits are relatively far away from the highest cross sections predicted by our model, but the highest values of the  SI cross section are within reach of the CDMS II search. It is also encouraging that the projected SuperCDMS detector would either discover or rule out our model in the first phase. It may thus happen that the first signal of our model would not be from the LHC, but rather from the detection of a dark matter particle. In such a case one would get a handle on both the mass and the coupling to the Higgs, which could be used to narrow down predictions for the LHC.

It would of course also be interesting to study the prospects for indirect searches for dark matter from annihilations in the sun or in the galaxy, but we leave this for the future.

%============================================================================== 
\section{Prospects for discovery at LHC}\label{Section:LHC}
%============================================================================== 

With improved naturalness and with a SM Higgs boson as heavy as $\sim 500$ GeV, the new states that are necessary to cut off the quadratic divergence in the Higgs mass need not be lighter than $\sim$ 1--2 TeV.  These new states need not carry color charge \cite{Chacko:2005pe} and will be very difficult to find at the LHC.  Even for new colored states this potentially pushes their mass well beyond the reach of the LHC \cite{Meade:2006dw}.  Thus, it may be that the only low energy states accessible by the LHC are the Higgs and the vectorlike charged and neutral leptons.  

A Higgs as heavy as $\sim 500$ GeV will be discovered in decays to four leptons with an integrated luminosity of less than $30$ fb$^{-1}$.  It is sufficiently broad\footnote{In the SM a Higgs of $500\,\gev$ has a width of $64\,\gev$; here the additional states available for its decay may appreciably alter the total width.} that it will even have its width measured to about 6\% accuracy after 300 fb$^{-1}$ \cite{ATLAS-TDR, CMS-TDR}.  Thus, if the lightest neutrino, $\chi$, is not observed in dark matter searches, after a few years of running at the LHC we may be left with the confusion of a heavy Higgs and no other new states to explain it, yet we know from EWPT that they must be there.

We now investigate the possibility of seeing the extra leptons at the LHC.  Analyses have been carried out in the past~\cite{Barger:1987re,Willenbrock:1985tj,Frampton:1992ik} for the case of an extra family of leptons. One major distinction from these previous works is the existence here of a heavy Higgs, which allows resonant production of the heavy leptons, greatly affecting the phenomenology.  Often analyses have assumed that the extra neutrinos are massless \cite{Barger:1987re}, or that the leptons are chiral so that for anomaly cancellation there are additional quarks which can affect production cross sections \cite{Willenbrock:1985tj}, or that the leptons are vector-like without Yukawa couplings and so do not couple to the Higgs. For a review of models with heavy leptons, see \cite{Frampton:1992ik}.

We define four possible classes of signatures and choose one particular point in parameter space for each of these signatures. These parameter points are chosen such that they are close to maximizing the signature under discussion and are therefore to be seen as best case scenarios. The chosen parameter points are displayed in Table~\ref{tab:points}, which shows the parameter values and the masses of the extra leptons. In addition we show the dark matter abundance, the contribution to $\Delta T$, and the change of the Higgs width for each of these points. The points are also displayed in the exclusion plots for dark matter searches, Figs.~\ref{DD_SI_plot} and \ref{DD_SD_plot}.

We have implemented the model in two multi-purpose matrix element generator packages: CalcHEP/CompHEP~\cite{Pukhov:2004ca,Boos:2004kh} and MadGraph/MadEvent~\cite{Stelzer:1994ta,Maltoni:2002qb,Alwall:2007st}. We have used combinations of these programs and the PYTHIA~\cite{Sjostrand:2006za} event generator to generate the hard subprocesses both for signal and background in the studies below. We have interfaced the parton level simulations to the PYTHIA and PGS~\cite{pgs} packages for full event simulation. The CalcHEP/CompHEP implementation has primarily been used to compute total cross sections and for interfacing to the micrOMEGAs program. We used the CTEQ6L1~\cite{Pumplin:2002vw} parametrization of the parton distribution functions in MadGraph/MadEvent\footnote{Note that all partonic level simulations in MadGraph/MadEvent were subject to the standard cuts: jet $p_T>20\,\gev$, photon $p_T>10\,\gev$; lepton $p_T>10\,\gev$, separation of jets, photons and leptons of $\Delta R=\sqrt{\Delta\eta^2+\Delta\phi^2}>0.4$; the jets, photons and leptons are all in the part of the detector with the best tracking, $|\eta|<2.5$.} and CTEQ5L~\cite{Lai:1999wy} in PYTHIA.

\subsection{LHC signals}

The possible production channels at our disposal are 
$q\bar q \to E^+ E^-$,
$q\bar q \to \nu_i \nu_j$,
$q\bar q' \to \nu_i E^\pm $, and 
$gg \to \nu_i\nu_j$ 
through an intermediate on-shell Higgs produced via a top loop. Because of the properties of the mass spectra, the decay channels are 
$\nu_3 \to E^\pm W^\mp$,
$\nu_3 \to \nu_{1,2} Z$, 
$\nu_2 \to \nu_1 Z$, and 
$E \to \nu_1 W$.  
The interesting experimental signatures are therefore based on ``short'' decay chains, the end products being weak bosons and missing energy.  This is unlike many of the cascades searched for as part of SUSY searches, since for us the weak bosons are always on shell.

Depending on the region of parameter space under study, we may concentrate on different signals.  The most interesting channels are where the gauge bosons decay leptonically leading to final states with varying numbers of leptons and missing energy.  We will not consider hadronic final states since they suffer from larger backgrounds, but the branching ratios are larger and these signals are worthy of further investigation.  It would seem that the states with the most leptons would be easiest to see but they require production of the more massive states higher up the decay chain and so suffer from small production cross sections.  

We study here the one, two and three lepton final states. In addition we consider the contribution of the new states to the total width of the Higgs boson, and we briefly consider the possibility of using the like-sign dilepton signal.

\begin{table}
\begin{center}
\begin{tabular}{|l|llll|lll|lll|}
\hline
Signal\ & $\lambda$ & $\lambda'$ & $M_N$ & $M_L$  
& $m_\chi$ & $m_2$  & $m_3$  
& $\Omega h^2$ & $\Delta T$ & $\Delta \Gamma_H$ \\
\hline
\hline
Higgs width   &  1.8  &  1.0  &  420  &  380  &  65  &  395  &  750  & 0.099 & 0.20 & 15.5 \\
One lepton    &  1.5  &  0.3  &  220  &  330  &  78  &  372  &  513  & 0.093 & 0.26 & 4.7 \\
Two leptons   &  1.7  &  0.6  &  440  &  280  &  83  &  310  &  666  & 0.091 & 0.17 & 9.3 \\
Three leptons &  1.4  &  0.8  &  100  &  280  &  78  &  300  &  478  & 0.096 & 0.16 & 7.2 \\
\hline
\end{tabular}
\caption{The benchmark points used in the text. For each point we show the four model parameters, the masses of the neutral leptons, the dark matter abundance, the contribution to the $T$-parameter, and the contribution to the total Higgs width. All dimensionful values are given in GeV. In all cases the Higgs mass is taken to be 500 GeV.
} \label{tab:points}
\end{center}
\end{table} 

\subsection{Higgs width}
Apart from the direct collider signatures, the effects of the new states can be searched for indirectly, for example through the contribution to the width of the Higgs boson. For Higgs bosons heavier than 200--230~GeV the width can be measured directly through the decay $H\to ZZ\to 4\ell$, and for the mass range we are interested in the width can be determined to roughly 6\% accuracy~\cite{ATLAS-TDR,CMS-TDR}.   The contribution to the width from a pair of heavy neutrinos is
\be\label{eq:higgswidth}
\Gamma_{h\to \nu_i \nu_j} = \frac{\left(m_H^2-(m_i+m_j)^2\right)^{3/2}  \left(m_H^2-(m_j-m_i)^2\right)^{1/2}}{8 \pi  m_H^3} U_{ij}^2.
\ee

\begin{figure}[t]
\begin{center}
\epsfig{file=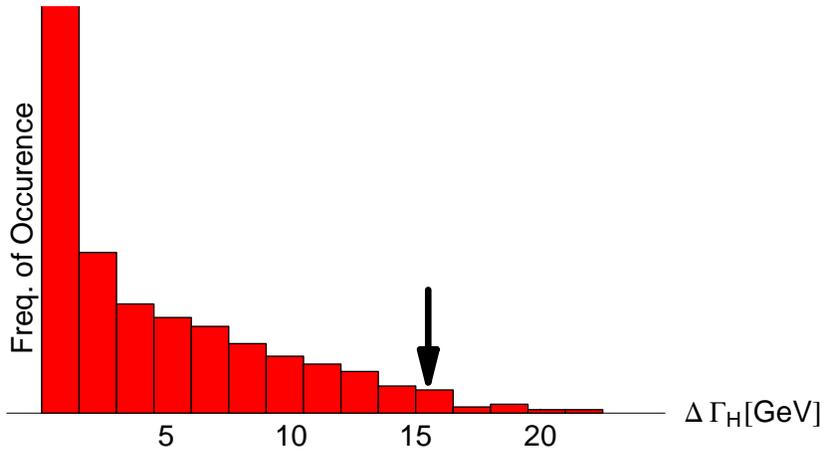, width=0.7\columnwidth}
\caption{Statistics of the correction to the Higgs width (in GeV) across the allowed parameter space. The arrow indicates the point given in Table~\protect\ref{tab:points}.\label{fig:higgswidth}}
\end{center}
\end{figure}

Thus if the new $H\nu_i\nu_j$ couplings give a large enough contribution to the total decay width of the Higgs boson, the shift compared to the expected SM width could be observed. In most of the allowed parameter space the shift is due to $H\rightarrow \nu_1 \nu_1$ and is a contribution to the invisible width.  We find that in about one third of the allowed parameter space this correction to the width is large enough to be observed (see Fig.~\ref{fig:higgswidth}). So although our four chosen parameter points all have relatively large contributions to the Higgs width we emphasis that this is not always the case.

\subsection{One lepton}

\begin{table}[t]
\begin{center}
\begin{tabular}{||c||c||c||c||c||}
\hline
Cuts  &    Signal   &$W^{*}\to \mu\nu$   & $WZ$    &      $t \bar t$ \\   
\hline
&$\sigma/$fb & $\sigma/$fb & $\sigma/$fb & $\sigma/$fb \\
\hline
\hline
No cuts                    & 19 &  $1.4 \times 10^7$  & 1000 &  320000   \\
 $N_{j}\le 1$, $N_\mu=1$   & 13 &  $1.3 \times 10^7$  &  590 &  4700   \\
 $H_{T}> 50\, \gev$        & 10 &  $2.3 \times 10^6$  &  350 &  4600    \\
 $M_{T}> 100\, \gev$       & 8.0 &  $4.7 \times 10^4$  &  190 &  1300   \\
 no tagged $b$             & 7.8 &  $4.4 \times 10^4$  &  190 &  1000  \\
\hline
\end{tabular}
\caption{Incremental effects of the cuts for the one lepton channel on the signal and the various SM backgrounds, in the case where the lepton is a muon.}
 \label{tab:1lept-cuts}
\end{center}
\end{table} 

The one lepton signal comes from the production channel $q\bar q' \to E^\pm \nu_1 \to W^\pm \nu_1 \nu_1$. The cross section is fairly decent:\ for the parameter point in Table~\ref{tab:points} we have $\sigma(E^\pm \nu_1)=$~110~fb, or if the ${\cal O}(\alpha_s)$ emission of one jet is included $\sigma(E^\pm \nu_1+\text{0,1 jets})=$~180~fb. This may at first sight seem promising as the signal is very simple:\ only a $W$ and missing energy, plus possibly jets from initial state radiation. For leptonic $W$ decays the signal is therefore one isolated lepton and missing energy. 
Unfortunately the missing transverse energy tends to cancel, so that the amount of missing energy is not large enough to use for triggering or kinematical cuts to suppress the background (it is peaked below 100 GeV). 

The main backgrounds to this process are Drell--Yan production, $W^{(*)}\to \ell \nu$, with $\ell$ the observed lepton or $W^{(*)}\to \tau \nu$ where the $\tau$ decays leptonically to the observed lepton;
$WZ$ production followed by $Z\to \nu \nu$ and $W\to \ell \nu$; and
$t\bar t$ production where there is a lepton either from the decay of a $B$ meson or from one of the $W$'s from the top quark decay.

We have made full simulations of the case where the lepton is a muon using MadEvent, PYTHIA and PGS, subject to the following cuts:
\begin{itemize}
\item exactly one isolated muon
\item $H_T>50$ GeV, where $H_T=\sum_\text{visible} |p_T|$ is the scalar sum of transverse momenta
\item $M_T>100$ GeV, where $M_{T}=\sqrt{2 \met p_T^\mu \left(1-\cos\Delta \phi\right)}$ is the transverse mass of the muon--missing energy system
\item a maximum of 1 jet in the central detector
\item no tagged $b$-jets in the event.
\end{itemize}
The effect of these cuts is summarized in Table~\ref{tab:1lept-cuts}.

These cuts suppress the signal by about $40 \%$ so that we expect about 780 signal events after 100 fb$^{-1}$. The $W^{(*)}\to\mu\nu$ background is the main problem because of its large production cross section, and we expect $4 \times 10^6$ events at the same luminosity. It does have a quite different kinematic structure than the signal, but imposing sharper kinematic cuts does not appear to help too much in improving the significance. We also expect about 20,000 $WZ$ events and 100,000 $t\bar t$ events. The former background is the most similar to the signal in terms of kinematical distributions, although the signal has a somewhat harder muon spectrum, which also leads to harder distributions in $H_T$ and $M_T$.
Another possibility is to concentrate on those events that have exactly one reconstructed jet, so that one may use it for kinematical variables. This does not help much, however, and it reduces the signal rate by about half.

The conclusion is that the one lepton signal is in practice unusable at the LHC. We expect similar conclusions when the lepton is an electron.
One reason for the dominance of the backgrounds is that the signal has a similar structure in that it only has a limited amount of missing energy. It is also hard to reduce the background since there are simply not enough kinematical variables to cut on.

\subsection{Two leptons}\label{dilepton}

Here the dominant production channels are $q\bar q, gg \to \nu_i\nu_1$, for $i=2,3$, with subsequent decay of $\nu_i\rightarrow \nu_1 Z$, and $q\bar q \to E^+ E^-$.  These have two lepton final states of $\nu_1\nu_1 l^+ l^-$ and $\nu_1\nu_1 l^+ l^- \nu_{SM}\bar\nu_{SM}$ respectively.  
Over some of parameter space the decay $H\to\nu_1\nu_2$ is open, which leads to a significant enhancement in the production cross section for this channel.  The signal was simulated at the parton level subject to some standard cuts and all events were then passed to PYTHIA and PGS to obtain fully simulated events.

\begin{figure}[t]
\begin{center}
\epsfig{file=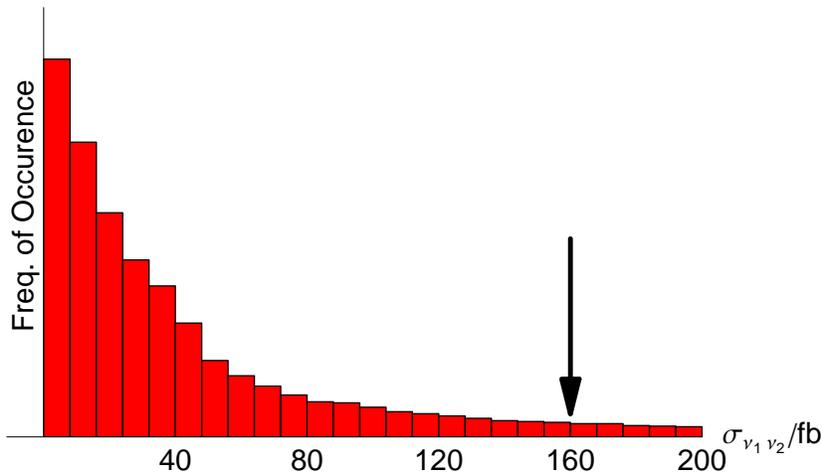, width=0.7\columnwidth}
\caption{Production cross section (in fb) for the two-lepton signal $\nu_1\nu_2$, sampled across the allowed parameter space. The point analyzed in the text has $\sigma\sim 160 $~fb  and is indicated by the arrow.}\label{2leptxsec}
\end{center}
\end{figure}

The distribution of production cross sections of $\nu_1 \nu_2$ sampled across the allowed parameter space is shown in Fig.~\ref{2leptxsec}.  For the example point given in Table~\ref{tab:points} the two production cross sections are $\sigma_{\nu_1\nu_2}\approx 160$~fb and  $\sigma_{E^+E^-}\approx 65$~fb.  Once the branching ratios of the $W$'s and $Z$'s into leptons (from now on we only consider muons) are taken into account the cross section into muons are $5$~fb and $0.7$~fb respectively. Because of the partonic level cuts on the leptons the vast majority of these events pass the PGS triggers, and so the effective cross section after triggers is comparable.  This dominant signal cross section ( $\sigma_{\nu_1\nu_2}$) would increase by a factor of two, modulo slight variations in triggering and detector efficiencies, if electrons were also included.

\begin{table}
\begin{center}
\begin{tabular}{||c||c||c||c||c||c||}
\hline
Cuts                                                                       &    Signal   &$WW$   & $ZZ$    &   $WZ$    &   $t \bar t$ \\   
\hline
&$\sigma/$fb & $\sigma/$fb & $\sigma/$fb & $\sigma/$fb & $\sigma/$fb\\
\hline
\hline
No cuts                                                                    &    5.7       &   920    &    140    &    290      &    8600 \\
$N_{j}<3$, $N_{Bj}=0$, $N_\mu=2$, $N_e=0$        &   4.5 	 &   490    &    80   &    86     &     870  \\
$|M_{\mu \mu}-m_Z|<10\,\gev$  			       &   4.1  	 &   60      &   61    &    56     &	110  \\
 $\Delta \phi_{\mu\mu}<2$         			       &   3.3 	 &   18    &   24   &   20     &	46	 \\
 $\Delta \phi_{Z\met}<2.5$        			       &   3.1 	 &   15     &   21   &   15     &	18	 \\
 $p_{T}^{\mu\mu}>100\,\gev$   			       &   2.3  	 &   0.79    &   10   &   5.5     &	4.5	 \\
 $\met>100\,\gev$                    			               &  1.3 	 &   0.43    &    8.9  &   3.2     &	1.3	 \\
 $|\eta_{\mu\mu}|<1$          				       &  1.3   	&   0.24     &    4.6  &   1.6     &	0.69	 \\
\hline
\end{tabular}
\caption{Incremental effects of the cuts for the dilepton case on the signal and the various SM backgrounds, in the case where the muons are leptons.}
 \label{tab:dilept-cuts}
\end{center}
\end{table}

For the two-lepton signal the dominant SM backgrounds are electroweak pair production of gauge bosons: $WW$ with both $W$'s decaying into muons (either directly or through taus), $ZZ$ with one $Z$ decaying into a muon pair and the other invisibly, and $WZ$ with the $Z$ again decaying into a muon pair, and the strong production of $t\bar{t}$ where both of the $W$'s decay into muons (either directly or through taus).  Note that 20\% of the time tau decays result in a muon but since these muons get only a fraction of the tau $p_T$ they are typically much softer than those produced promptly.  If the taus are produced from $Z$ decay the resulting muons are unlikely to reconstruct to the $Z$ peak and so constitute only a small background, but for the case of $t\bar{t}$, and to a lesser extent $WW$, the tau contribution is sizable.  For the case of $WZ$ the contribution from all leptonic decay channels of the $W$, with the $Z$ decaying to muons, must be included.   In addition to these backgrounds there are others where one $Z$ decays into muons and the other $W$ or $Z$ has hadronic decays, but due to the extra jet activity and the lack of missing energy these will be completely removed by the analysis cuts.  Once the relevant branching fractions are taken into account the backgrounds have production cross sections of: $\sigma_{WW}\sim 0.9$~pb, $\sigma_{ZZ}\sim 0.1$~pb, $\sigma_{WZ}\sim 0.3$~pb, $\sigma_{t\bar{t}}\sim 9$~pb.  The backgrounds were generated using PYTHIA and then passed to PGS. 

\begin{figure}[t]
\begin{center}
\epsfig{file=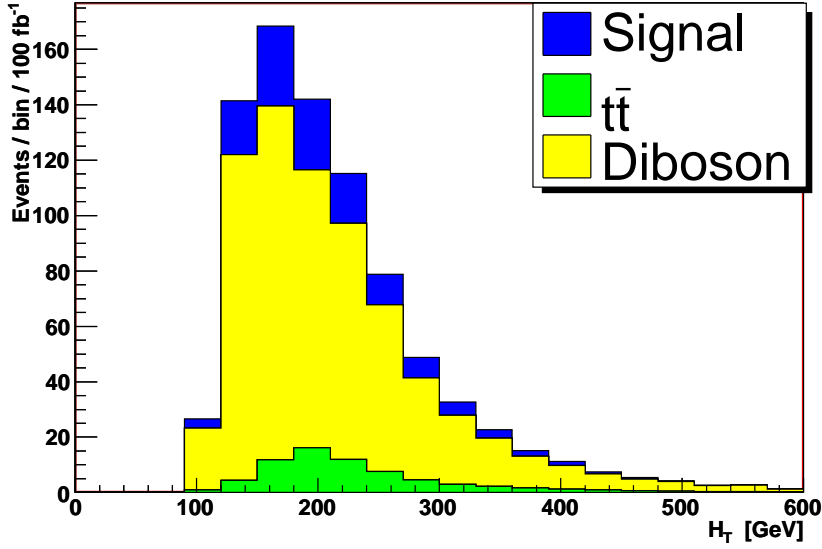, width=0.7\columnwidth, 
bbllx=13 , bblly=8 , bburx=522 , bbury=376 , clip=}
\caption{Distribution of $H_T=\sum_\text{visible} |p_T|$ for the 2 lepton (muon only) signal after 100 fb$^{-1}$.  The green is $t\bar{t}$ background, yellow is the diboson background  and blue is the signal.}\label{HT2lep}
\end{center}
\end{figure}

The signal is fairly clean, consisting of 2 muons and missing energy ($\met$) whereas many of the backgrounds have more particles in the final state.  For the dominant signal channel the 2 muons will reconstruct an on-shell $Z$.  This $Z$ comes from the cascade of a heavy state down to $\nu_1$ and because the splitting of $\nu_2$ and $\nu_1$ is always large the $Z$ will be produced with significant $p_T$, thus the $p_T$ of the dimuon pair, and consequently $\met$, will be large.  The large $p_T$ of the $Z$ also means that the two muons will be somewhat collimated.  These observations lead to a series of cuts that have little effect on the signal but greatly reduce the background.  They are,
\begin{itemize}
\item Exactly 2 hard muons, no electrons, no $b$-jets and fewer than 3 jets in the event.
\item The dimuon invariant mass is close to $m_Z$: $80\,\gev<M_{\mu\mu}<100\,\gev$\item Muons are somewhat collimated: $\Delta  \phi _{\mu\mu}<2$.
\item The dimuon is back-to-back with the missing energy, limits the amount of other activity in the event: $\Delta  \phi _{Z\slashed{E}_T}>2.5$
\item Dimuon has large transverse momentum: $p_T^{\mu\mu}> 100\,\gev$
\item Large missing energy: $\met>100\,\gev$.
\item The dimuon is central: $|\eta_{\mu\mu}|<1$
\end{itemize}
The efficiency of these cuts to lower the background while preserving the signal is shown in Table~\ref{tab:dilept-cuts}.  A complete analysis involving other possible backgrounds, full detector effects etc. will doubtless change these numbers somewhat.  It is worth noting that, for the dominant backgrounds, one will not have to rely on a simulation; it will be possible to directly measure these backgrounds in the region of interest.  The $t\bar{t}$ background can be measured in the semi-leptonic channel while the $ZZ$ and $WZ$ backgrounds will be measured in the fully leptonic channel.  

From Table~\ref{tab:dilept-cuts} we see that after an integrated luminosity of 100 fb$^{-1}$ the significance ($S/\sqrt{B}$) is roughly 4.8.  We have checked that in the case of the signal the efficiencies are similar for the $e^+ e^- + \met$ final state so the inclusion of the electron final state could raise the significance by a factor of $\sim \sqrt{2}$. Despite the significance being fairly large in the two lepton channel the signal and background are similarly distributed making it hard to easily separate them.  In Fig.~\ref{HT2lep} we plot the distribution of $H_T$ after all the cuts are imposed for the background and signal.  As can be seen from Fig.~\ref{2leptxsec} the point analyzed here is optimistic, for a more average point the signal could be substantially reduced making discovery in this channel impossible.

\subsection{Three leptons}

\begin{figure}[t]
\begin{center}
\epsfig{file=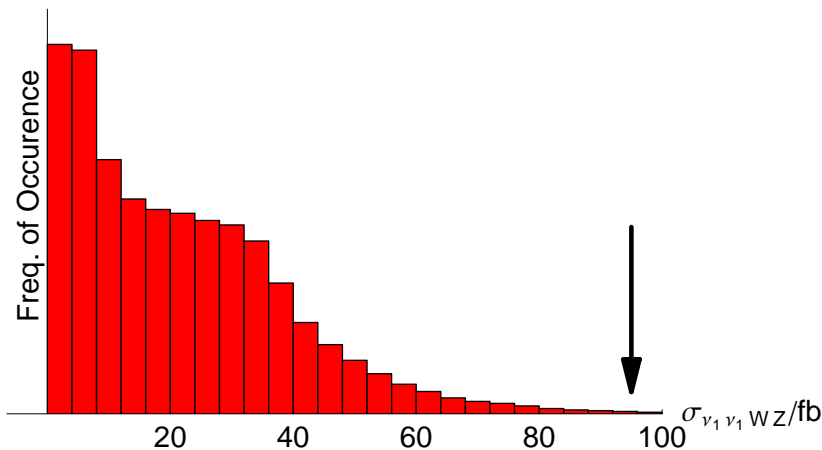, width=0.7\columnwidth}
\caption{Production cross section in fb for the three-lepton signal $\nu_1\nu_1 W^\pm Z$, sampled across the allowed parameter space. The point analyzed in the text has $\sigma\sim 95 $~fb  and is indicated by the arrow.}\label{fig:xsct-WZNN}
\end{center}
\end{figure}

We now focus on the associated production of a heavy charged lepton with a heavy neutrino  $q\bar q' \to E^{\pm} \nu_{2,3} \to W^{\pm}Z \nu_{1}\nu_{1}$. Since it is a Drell--Yan process with quite heavy final states ($\sqrt{\hat s}\gtrsim 500\,\gev$) the production cross section is smaller than in the previous cases, typically of the order of 10--100~fb.
In particular the distribution of the cross section $\sigma(p\,p\to W^{\pm}Z\nu_{1}\nu_{1})$ over the allowed parameter space is shown in Fig.~\ref{fig:xsct-WZNN}. 
The presence of both a $W$ and a $Z$ leads to richer final states and this might help in reducing the SM backgrounds.
The subsequent decay of the gauge bosons will give final states with 3 leptons + $\met$, 2 leptons + 2 jets + $\met$ or 1 lepton (+ 2 jets) + $\met$.
We do not consider here the case of one lepton and missing transverse energy since it has already been studied in the previous sections, while the 2 leptons + 2 jets + $\met$ is not expected to give better results and it is left to further study. 

The remaining case with 3 leptons is a standard SUSY search for associated  $\tilde\chi_1^\pm \tilde\chi_2^0$ production. Various studies have been performed for the LHC \cite{Baer:1994nr,Muanza:1997,CMS-TDR,ATLAS-TDR} and preliminary bounds from CDF and D0 are available \cite{Canepa:2006eu,Bortoletto:2005jq}. However the smallness of the cross-section makes our case beyond Tevatron reach. Moreover the LHC studies mentioned above have been performed in the context of mSUGRA and they focus on regions of parameter space where $m_{\tilde\chi_{2}}-m_{\tilde\chi_{1}}<m_{Z}$, giving rise to an off-shell $Z$ decaying into a pair of leptons. This is not the case of our model where the constraints coming from $\Delta T$ and from the DM abundance force the $\nu_{2}$--$\nu_{1}$ splitting to always be larger than the $Z$ mass. We are aware of only one analysis with on-shell $Z$ \cite{CMS-TDR}:\ this is an inclusive analysis and requires much higher production cross-sections than the ones available in our model. Motivated by these considerations we perform a brief study of the trilepton signal in the following.

We use the point shown in Table~\ref{tab:points}, giving rise to a production cross section of $\sigma_{WZ\nu_{1}\nu_{1}} =95.4$~fb, as shown by the arrow in Fig.~\ref{fig:xsct-WZNN}. Once the branching ratios of the weak gauge bosons to leptons (here we will consider only muons and electrons and leptonic decays of $\tau$'s) are taken into account we obtain a final cross section of 2.25~fb.
Given the smallness of the cross-section, assessing the observability of this signal will ultimately require a more careful study of the backgrounds with proper detector simulation. Here we will focus mainly on the most important reducible and irreducible backgrounds. 

The main SM backgrounds are $WZ$ diboson production with both the $Z$ and the $W$ decaying leptonically, $t\bar t$ production with both $t$'s decaying leptonically and with a third lepton coming from a semileptonic $b$ decay or from a jet fake, and associated production of $Z$ with heavy quarks decaying semileptonically.  
The signal was simulated using the same combination of MadGraph/MadEvent, Pythia and PGS, with the same selection cuts and triggering criteria as used above.
The $WZ$ and $t\bar t$ backgrounds were simulated using Pythia and passed again through PGS for event reconstruction, while for the $Z \bar b b$ and $Z \bar c c $ ones the same software used for the signal has been utilized.

\begin{table}
\begin{center}
\begin{tabular}{||c||c|c|c|c||}
\hline
Cuts & {Signal} & {$W\,Z$} & {$t \bar t$} & {$b \bar b Z,\, c \bar c Z$} \\   
\hline
& $\sigma/fb$ & $\sigma/fb$ & $\sigma/fb$ & $\sigma/fb$ \\
\hline
\hline
None 								                  & 1.8 	& 700 			   & $3.6 \times \, 10^{4}$  & $2.3\times\, 10^{4}$ \\
 $N_{jet}<3$, $N_{bjet}=0$, $N_{leptons}>2$  & 0.74  	& 190  			   & $1.6\times\, 10^{3}$  & 640\\
 $|M_{\ell \ell}-m_Z|<10\,\gev$   		      & 0.55 	& 170			   & 53 		         & 370\\
 $\Delta R_{\ell\ell}<2.3$ 				      & 0.43 	& 60 			      & 18 		         & 100\\
 $M_{T}>150\,\gev$   					         & 0.15 	& $3.6\times \, 10^{-2}$	& 0.18		      & $< 1.6\times\, 10^{-2}$\\
 $M_{\ell j}<60\,\gev$ 					         & 0.12 	& $3.3\times \, 10^{-2}$	& 0.09		      & $< 1.6\times\, 10^{-2}$\\
\hline
\end{tabular}
\caption{Incremental effects of the cuts for the trilepton case on the signal and the various SM backgrounds.}
 \label{tab:trilept-cuts}
\end{center}
\end{table}

\begin{figure}[t]
\begin{center}
\epsfig{file=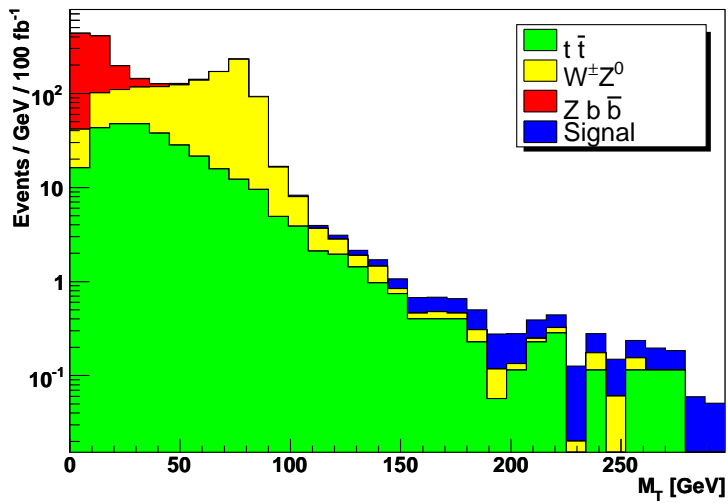, width=0.7\columnwidth}
\caption{Transverse mass distribution from third lepton and missing transverse energy, for the trilepton signal, $WZ$ and $t \bar t$ backgrounds after cuts on the invariant mass of the dilepton pair, where the leptons include muons and electrons.}\label{fig:jacobianw}
\end{center}
\end{figure}

Since part of the backgrounds was due to non-isolated leptons and/or jet fakes which are sensitive to the jet reconstruction and the detector simulation, we performed the analysis using both $k_T$ and cone algorithms.
The cross sections for $WZ$, $t \bar t$ and $Z Q \bar Q$ with vector bosons decaying leptonically are 0.53~pb, 34~pb and 20~pb respectively.
This includes also the $\tau$ channels, with $\tau$'s decaying into leptons.
The signal involves an opposite sign (OS) same flavor (SF) lepton pair with invariant mass around $m_{Z}$ and a third (unpaired) lepton together with large $\met$. The large $\met$ comes from the two $\nu_1$ and from the neutrino coming from W decay. Hence the transverse mass built from the unpaired lepton and the $\met$ should not exhibit a Jacobian peak around any physical mass. This is not the case for the $WZ$ background where the same kinematic variable shows a strong Jacobian peak at $m_{W}$ as shown in Fig.~\ref{fig:jacobianw}.

The list of cuts we use is then:
\begin{itemize}
\item Muon isolation cut: the muons considered in the analysis have to be isolated ($<10\,\gev$ of energy deposited in a 30-degree cone around the track).
\item Dilepton Z peak: two OS SF isolated leptons with invariant mass $M_{\ell\ell}$ in the range $80\,\gev<M_{\ell\ell}<100\,\gev$.
\item High transverse mass: $M_{T}=\sqrt{2 \met p_{T3} \left[1-\cos\left(\Delta \phi\right)\right]}>150\,\gev$, where $p_{T3}$ is the transverse momentum of the third (unpaired) isolated lepton and $\Delta \phi $ is the azimuthal angle between the lepton and the missing energy vector.
\item no tagged $b$-jets.
\item Small hadronic activity: less than 3 jets in the event.
\item Dilepton separation: the OS SF leptons should have $\Delta R<2.3$.
\item the minimum of the invariant masses of the same sign leptons or the unpaired lepton with the highest-$p_{T}$ jet should not be greater than $60\,\gev$.
\end{itemize}
The last cut is motivated by strategies to measure the top mass in the leptonic channel which use the kinematic variable \cite{ATLAS-TDR}:
\beq
m_{t}^{2}=m_{W}^{2}+2\frac{\langle m_{lb}^{2} \rangle}{1-\langle \cos (\theta_{lb})\rangle}
\eeq
where $m_{lb}$ is the invariant mass of the lepton and the $b$-jet from the same side of the event and $\theta_{lb}$ is their opening angle in the $W$ rest frame.
For our case one has also to keep in mind that $t \bar t$ is a background for the trilepton signal when one of the leptons is faked by a jet. This cut has the effect of reducing the $t\bar t$ background by an additional factor of a few.

\begin{figure}[t]
\begin{center}
\epsfig{file=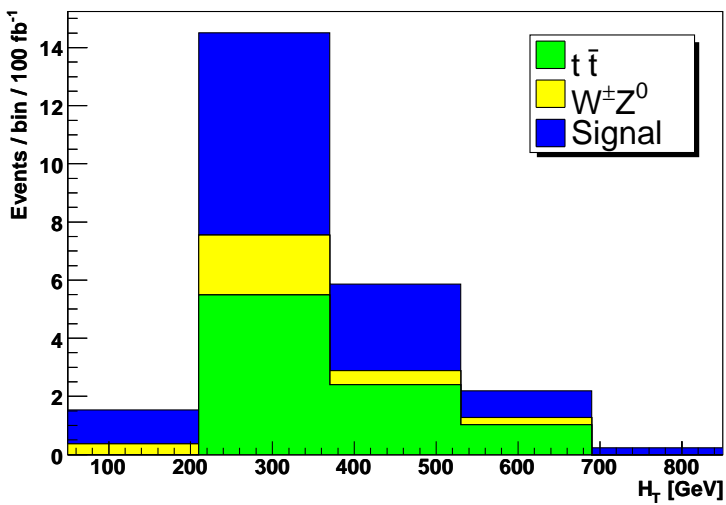, width=0.7\columnwidth}
\caption{Number of events after $100\,fb^{-1}$ for signal and backgrounds as a function of the scalar sum of visible energy $H_{T}$.}\label{fig:trilept-ht}
\end{center}
\end{figure}

Moreover the leptons in the signal tend to be isolated and there is no heavy flavor involved in the process, so isolation cuts and $b$-jet vetoes can be further used to reduce the background, especially the $t\bar t$ one.
Finally the $Z$ in the signal will generically be more boosted than in the $SM$ diboson case, so the opening angle of the dilepton pair will be smaller for the signal than for the background.

The cut efficiencies for signal and backgrounds are shown in Table~\ref{tab:trilept-cuts}.
The backgrounds can be drastically reduced to a rate comparable to the signal. However the cuts have the effect of also reducing the signal by more than an order of magnitude. The cross sections for signal and backgrounds are also given in Table~\ref{tab:trilept-cuts}, and the differential distribution as a function of $H_{T}$ for the event rate after 100~fb$^{-1}$ is shown in Fig.~\ref{fig:trilept-ht}. 
By using these numbers one gets a $S/\sqrt{B}\sim 3.5$, an evidence for the trilepton signal after 100~fb$^{-1}$ and $S/\sqrt{B}\sim 6$ after 300~fb$^{-1}$. However one has to keep in mind that the above analysis is not fully realistic, and the above numbers will likely change with improvements in detection simulation and reconstruction, and if additional sources of background are found to be important. Moreover as shown in Fig.~\ref{fig:xsct-WZNN} the benchmark point used in the analysis has a rather large cross section, while the bulk of the allowed parameter space has smaller cross sections (by a factor of $2-3$).

\subsection{Like-sign dileptons}

The Lagrangian of (\ref{eqn:lagrangian}) contains a Majorana mass term. This allows for the production of like-sign heavy leptons through two processes:\ $W$ boson fusion to two charged heavy leptons, and  production through an $s$-channel $W$ giving two like-sign charged heavy leptons with an off-shell intermediate heavy neutrino. The two heavy charged leptons are produced together with two jets, and subsequently decay into like-sign muons and electrons. Boson fusion primarily leads to forward jets, while the other process leads to more central jets. In both cases the decays of the charged leptons lead to heavy neutrinos and SM neutrinos in the final state potentially resulting in large missing energy. Similar signals have been studied in \eg, Refs.~\cite{Han:2006ip,Alwall:2007ed}, but these signals did not involve missing energy. 

The production cross section $\sigma(E^\pm E^\pm j j)$ for the case of the one-lepton parameter point in Table~\ref{tab:points} is roughly 3.7 fb when requiring $|\eta_j|<4.5$, and 3.4 fb when requiring $|\eta_j|<2$. The forward jet cross section is thus appreciably smaller than the central jet cross section. Requiring the final state to consist of muons and electrons leads to cross sections of 0.2 fb for central jets and 0.02 fb for forward jets. 

The backgrounds are numerous, consisting of, \eg, single weak boson production with three jets, one of which fakes a lepton; diboson production with two jets; $W^\pm W^\pm W^\mp$, where the opposite sign $W$ decays hadronically; and $t\bar t$. In many cases one charged lepton must escape undetected or a jet must fake a lepton. Many of these backgrounds have sizable cross sections, but the kinematics may be different from the signal. In particular, the missing energy characteristics may be sufficiently different to allow separation of signal and background. We leave this analysis for the future.

%============================================================================== 
\section{Conclusions}
%============================================================================== 

If the Standard Model describes electroweak symmetry breaking to high
precision, then the measurements of the $S$ and $T$ parameters imply
that the Higgs boson is lighter than about 150 GeV.  Naturalness would
then imply new physics, typically involving colored states, at or below
500 GeV. This argument, that LHC will probe the physics that completes
the Standard Model into a more natural theory, is badly flawed. The
$S$ and $T$ parameters have only a weak logarithmic dependence on the
Higgs mass, and if there is additional physics containing states with
electroweak quantum numbers with masses of several hundred GeV, then
it is perfectly natural that these states make similar contributions to
$S$ and $T$ as the Higgs, so that the Higgs mass becomes
unconstrained.  Indeed, if the UV completion of the Standard Model
involves some sort of new strong interactions below 10 TeV,  then a
light Higgs is strongly disfavored, since it requires fine tuning to
keep it light.  Further motivation for these electroweak states, which
need not be colored, is that they may account for the observed dark
matter.   

In this paper we have focused on the critical question of whether  LHC can
probe these new electroweak states that improve naturalness, by
investigating a very simple model that augments the Standard Model
with a vector lepton doublet and a Majorana singlet.  These new states
carry some new parity so that they do not mix with the known leptons;
this reduces the parameter set
and the lightest new state, $\chi$, is stable.  As well as two
mass parameters, which we assume are of order the electroweak scale,
the model contains two new Yukawa couplings; for
simplicity a possible extra phase is set to zero.  By studying the
contribution of these new states to $S$ and $T$, we
verify that there is a wide range of parameters that lead to a heavy
Higgs boson, for example in the range of 400 to 600 GeV, as
illustrated in Fig.~\ref{Tplot}.   The new Yukawa couplings must be
of order unity, and, together with the heavy Higgs, this brings up the
issue of naturalness and perturbativity. For values of the new Yukawa
couplings less than 2 (which is the bound we imposed on our parameter space)
the improved naturalness of the Standard Model with a heavy Higgs is
maintained. Furthermore, for Higgs bosons as heavy as 500 GeV, values
of these Yukawa couplings near 2 lead to the Higgs quartic coupling remaining
perturbative to higher energies than in the Standard Model, as shown
in Fig.~\ref{Hcut}.    

Within a sub-space of the parameter range that leads to a heavy Higgs, 
$\chi$ accounts for the observed amount of dark matter in the
universe, as illustrated in Fig.~\ref{DMplot}. The allowed region in
Fig.~\ref{DMplot} appears quite small; but this reflects our precise
knowledge of the dark matter density and not any need for fine
tuning. Indeed, the constraint from dark matter greatly reduces the 
parameter space that needs to be experimentally explored at LHC, 
leading to the characteristic spectrum of the new electroweak states 
shown in Fig.~\ref{fig:spectrum}.  The dark matter particle $\chi$ has
a mass in the range 50--170 GeV; direct detection of dark matter will
probe part of the allowed parameter space if the reach on the cross
section is improved by one order of magnitude, and the entire
parameter space with two orders of magnitude improvement, as shown in
Figs.~\ref{DD_SI_plot} and \ref{DD_SD_plot}.  

If our model is correct, the first discovery at LHC will be the heavy
Higgs, via the usual $WW$ and $ZZ$ modes.  The measurement of $m_H$ will
narrow the value of $\Delta T$ needed from the new electroweak states. If
dark matter is directly detected, the additional constraints from the
abundance, cross section and mass of $\chi$ will lead to measurements
of all 4 new parameters of the model, although with important uncertainties
from the local halo density.  This will determine whether or not
further signals will be observed at LHC.  Observation of further
signals, together with the absence of other new physics, would
strengthen the evidence for this theory.  

For a Higgs boson mass of 500 GeV, approximately a third of the presently
allowed parameter space will lead to a measurable increase of the
total Higgs width by 6\% or more.  The remaining LHC signals that we
have studied are multi-lepton events arising from the production of the
new electroweak states, followed by their decay via $W$ and $Z$ bosons to
$\chi$.  The pair production cross sections for these states are
typically in the range of 10--100 fb, and the price of leptonic branching
ratios must be paid to obtain events with 1, 2, or 3 isolated charged
leptons.  Background events from $WW$, $WZ$, $ZZ$ and $\bar{t}t$ production
have much larger rates, so that great care is necessary in devising cuts that
are efficient at removing background while preserving the signal. 

We have studied the multilepton signals in some detail for
benchmark points in parameter space that are close to maximizing the
relevant production cross sections. The signal and backgrounds  were simulated using
MadGraph/MadEvent and Pythia,  and passed through PGS for event
reconstruction. In
the case of single lepton events, for all points in parameter space,
we find that there are insufficient measured variables to allow a
signal to emerge from the large background.   
The series of cuts applied to dilepton and trilepton
event samples are shown in Tables \ref{tab:dilept-cuts} and
\ref{tab:trilept-cuts},  respectively, together
with the resulting decreases in both signal and background cross
sections. For dileptons, the cuts deplete the $WW$ and $\bar{t} t$
backgrounds by of order $10^4$, while losing a factor 4--5 in
signal. For trileptons, backgrounds are removed at the factor $10^5$
level, for an order of magnitude cost in signal. These high rejection
levels are a result of the cumulative effect of several cuts, so that
the surviving backgrounds have similar distributions to the signal, as
can be seen in the distributions for the scalar sum of visible $p_T$
shown in Figs.~\ref{HT2lep} and \ref{fig:trilept-ht}.   For an
integrated luminosity of 100 fb$^{-1}$, we estimate $S/\sqrt{B}$ of
4.8 and 3.5 for the di- and trilepton cases respectively; but these
numbers are likely to change with improvements in simulations and
better understanding of backgrounds. In much of the allowed parameter
space the signal cross section is a factor 2 to 10 smaller.

Although our model is particularly simple, there are many variants 
that have a heavy Higgs boson allowing an improved naturalness so that 
new colored states are not expected below 2 TeV.  Discovery of the 
new electroweak states at LHC will be difficult, with multilepton 
signals similar to the minimal model studied here.  Nevertheless, 
by itself the discovery of a heavy Higgs would indicate a new direction, 
and the presence of the electroweak states could be confirmed by direct 
detection of dark matter. 

\begin{acknowledgments}
We thank Jeffrey Filippini, who has been very helpful, answering our many questions and providing the exclusion limits for dark matter from the Dark Matter Limit Plot Generator. We also thank Johan Alwall for help with MadGraph and MadEvent. PJF and MP thank the Aspen Center for Theoretical Physics for hospitality. This research was supported in part by the Director, Office of Science, Office of High Energy and Nuclear Physics, Division of High Energy Physics of the US Department of Energy under contract DE-AC02-05CH11231, in part by US Department of Energy contracts DE-FG02-04ER41319 and DE-FG02-04ER41298, in part by National Science Foundation Grant PHY-04-57315, and in part by The Swedish Research Council.
\end{acknowledgments}

\appendix

%============================================================================== 
\section{The heavy neutrino mass eigenvalues and eigenvectors}\label{Appendix:Eigenvals}
%============================================================================== 

In the numerical calculations presented we diagonalize the heavy neutrino mass matrix numerically, but for completeness we here give analytical expressions for the masses.
The eigenvalues can be written as
\be
m_i=\frac{M_N}{3}+2 r^{1/3}\times \left\{ \cos \frac{\theta}{3},\ \cos \frac{\theta+2\pi}{3},\ \cos \frac{\theta+4\pi}{3}\right\}
\ee 
where 
\begin{align}
r&=\frac{1}{3\sqrt{3}}\left[(\lambda^2+\lambda'^2)v^2+M_L^2+\frac{M_N^2}{3}\right]^{3/2}
\\
\cos\theta&=\frac{1}{2r}\left[ \frac{1}{3}(\lambda^2+\lambda'^2)v^2 M_N-\frac{2}{3}M_N(M_L^2-\frac{1}{9}M_N^2) +2 M_L\lambda \lambda' v^2 \right].
\end{align}
For eigenvalue $m_i$, the corresponding eigenvector $v_i$ is
\be
\frac{1}{{\cal N}_i}
\begin{pmatrix}
m_i^2 -M_L^2\cr
(m_i \lambda +M_L \lambda')v \cr
(m_i \lambda' +M_L \lambda)v 
\end{pmatrix}
\ee
where the normalization is ${\cal N}_i^2=(M_L^2-m_i^2)^2+((m_i \lambda +M_L \lambda')^2+(m_i \lambda' +M_L \lambda)^2)v^2$.

%=============================================================================
\section{WIMP annihilation cross sections}\label{Appendix:DM}
%=============================================================================

\subsection{$\chi\chi\to f\bar f$}
The dark matter particle is the lightest neutrino $\nu_1$, which we shall refer to as $\chi$.
The coupling of $\chi$ to the $Z$ boson is given by $\frac{i g}{2c_W}V_{11} \gamma^\mu \gamma_5$ and the coupling of the fermion $f$ of mass $m_f$ to the $Z$ is $\frac{i g}{2c_W}\gamma^\mu (C_V-C_A \gamma_5)$. 
The thermally averaged annihilation cross section can be expanded in powers of the relative velocity $\vr=2 v$,
\be
\vev{\sigma \vr} = a+b\vev{\vr^2} = a+b \frac{6T}{m_\chi},
\ee
by using the non-relativistic expressions for the Mandelstam variables
\begin{align}
s=&4m_\chi^2(1+v^2) \\
t=&m_f^2-m_\chi^2+ 2 m_\chi^2\sqrt{1-\z} \; \cos \theta \; v-2m_\chi^2 v^2,
\end{align}
where $v$ is the CM velocity and $\theta$ the CM scattering angle. 
One finds (neglecting Higgs exchange)
\begin{align}
a_{f\bar f} =&  \frac{2\pi\alpha^2}{s_W^4 c_W^4}  \sqrt{1-\frac{m_f^2}{m_\chi^2}} \;  \dfrac{|V_{11}|^2}{(4m_\chi^2-m_Z^2)^2+m_Z^2\Gamma_Z^2} \; C_A^2 m_f^2 \left( 1- \frac{4 m_\chi^2}{m_Z^2} \right)^2
\\
b_{f\bar f} =& \frac{2\pi\alpha^2}{3s_W^4 c_W^4} \sqrt{1-\dfrac{m_f^2}{m_\chi^2}} \;  \dfrac{m_\chi^2  |V_{11}|^2}{(4m_\chi^2-m_Z^2)^2+m_Z^2\Gamma_Z^2} \\
&\times \left\{
C_V^2 \left[1+\frac{1}{2} \z \right] 
+
C_A^2 \left[1 - \frac{1}{8} \z \left(17-\frac{3}{1-\sfrac{m_f^2}{m_\chi^2}}\right)\right]
\right. \nonumber \\
&-\left. 3 C_A^2 \frac{m_f^2}{m_\chi^2} \left[
 \frac{1}{1-\sfrac{m_f^2}{m_\chi^2}} \left(  \frac{m_f^2}{ m_Z^2}- 2\frac{m_\chi^4}{ m_Z^4} \left(2-\z\right)\right) +
\frac{2m_\chi^2}{m_Z^4} \dfrac{(4m_\chi^2-m_Z^2)^3}{(4m_\chi^2-m_Z^2)^2+m_Z^2\Gamma_Z^2}
\right]
\vphantom{\frac{1}{1-\sfrac{m_f^2}{m_\chi^2}}} % To get right size }
\right\}.\nonumber
\end{align}
To obtain the full cross section this must be summed over all fermion species allowed by the available phase space.

\subsection{$\chi \chi \to W^+ W^-$}
Neglecting Higgs exchange we have two diagrams with exchange of the charged lepton in the $t$- and $u$-channels and one diagram with exchange of a $Z$ in the $s$-channel. The $W^\pm E^\mp \chi$ vertex is 
$\frac{i g}{\sqrt{2}} \gamma^\mu(g_V-g_A \gamma_5)$ where $g_{V,A}=g_{V,A}^{i=1}$. 
We obtain
\begin{align}
a_{WW} =&  \frac{\pi\alpha^2(g_V^2+g_A^2)^2}{s_W^4}   \left(1-\frac{m_W^2}{m_\chi^2}\right)^{3/2} \;  \left[\dfrac{m_\chi}{m_\chi^2+M_L^2-m_W^2}\right]^2.
\end{align}
The coefficient $b_{WW}$ is lengthy and not very illuminating, so we omit its full expression here. 

\subsection{$\chi \chi \to Z Z$}
Again neglecting Higgs exchange, we now have exchange of the three neutrinos in the $t$- and $u$-channels. The $Z \nu_i \chi$ vertex is $\frac{i g}{2c_W} V_{i1} \gamma^\mu \gamma_5$. This gives
\begin{align}
a_{ZZ} =&  \frac{\pi\alpha^2}{8c_W^4}  \left(1-\frac{m_Z^2}{m_\chi^2}\right)^{3/2} \;  \left[\sum_{i=1}^3  V_{i1}^2 \dfrac{m_\chi}{m_\chi^2+m_i^2-m_Z^2}\right]^2.
\end{align}
We omit the explicit expression for $b_{ZZ}$, which is, again, somewhat lengthy.

%%%%%%%%%%%%%%%%%%%%%%%%%%%%%%%%%%%
\bibliography{EFHPP}
\bibliographystyle{JHEP}
%%%%%%%%%%%%%%%%%%%%%%%%%%%%%%%%%%%

\end{document}